\documentclass[fleqn,usenatbib]{mnras}

\usepackage{newtxtext,newtxmath}

\usepackage[T1]{fontenc}
\usepackage{ae,aecompl}

\usepackage{amsfonts}
\usepackage{amsmath}
\usepackage{amssymb}
\usepackage[ngerman, british]{babel}
\usepackage{booktabs}
\usepackage{color}
\usepackage{enumitem}
\usepackage{graphicx}
\usepackage[utf8]{inputenc}
\usepackage[]{tabularx}

\newcommand{\mum}{\,\mu\hbox{m}}

\newcommand{\Lsun}{L_\odot}

\title[Modelling 49~Ceti]{A multi-wavelength study of the debris disc around 49~Cet\thanks{Based on observations collected at the European Southern Observatory under ESO programmes 198.C-0209(N) and 097.C-0747(A)}}

\author[N. Pawellek et al.]{Nicole Pawellek$^{1,2}$ \thanks{E-mail: pawellek@mpia.de}, 
Attila Mo\'or$^{2}$,
Julien Milli$^{3}$,
\'Agnes K\'osp\'al$^{1,2}$,
Johan Olofsson$^{1,4,5}$,
\newauthor
P\'eter \'Abrah\'am$^{2}$,
Miriam Keppler$^{1}$,
Quentin Kral$^{6}$,
Adriana Pohl$^{1}$,
\newauthor
Jean-Charles Augereau$^7$
Anthony Boccaletti$^6$,
Ga\"{e}l Chauvin$^8$,
\'Elodie Choquet$^{9}$ 
\newauthor
Natalia Engler$^{10}$,
Thomas Henning$^{1}$, 
Maud Langlois$^{11}$
Eve J. Lee$^{12}$,
\newauthor
Fran\c{c}ois M\'enard$^{7}$
Philippe Th\'ebault$^{6}$,
Alice Zurlo$^{13, 14, 15}$
\\
$^{1}$ Max-Planck-Institut f\"ur Astronomie, K\"onigstuhl 17, 69117 Heidelberg, Germany\\
$^{2}$ Konkoly Observatory, Research Centre for Astronomy and Earth Sciences, Hungarian Academy of Sciences, \\ Konkoly-Thege Mikl\'os \'ut 15-17, H-1121 Budapest, Hungary\\
$^{3}$European Southern Observatory (ESO), Alonso de C\'ordova 3107, Vitacura, Casilla 19001, Santiago, Chile\\
$^{4}$ Instituto  de  F\'isica  y  Astronom\'ia,  Facultad  de  Ciencias,  Universidad de Valpara\'iso, Av. Gran Breta\~{n}a 1111, Playa Ancha,\\ Valpara\'iso, Chile\\
$^{5}$  N\'ucleo Milenio Formaci\'on Planetaria–NPF, Universidad de Valpara\'iso, Av. Gran Breta\~{n}a 1111, Valpara\'iso, Chile\\ 
$^6$ LESIA, Observatoire de Paris, Universit\'e PSL, CNRS, Sarbonne Universit\'e, Univ. Paris Diderot, \\ Sorbonne Paris Cit\'e, 5 Place Jules Janssen, 92195 Meudon, France\\   
$^{7}$ Univ. Grenoble Alpes, CNRS, IPAG, 38000 Grenoble, France\\
$^8$ International Franco-Chilean Laboratory of Astronomy, CNRS/INSU UMI3386, Department of Astronomy, \\ University of Chile, Casilla 36-D, Santiago, Chile\\
$^{9}$ Aix Marseille Univ., CNRS CNES, LAM, P\^{o}le de l’\'Etoile Site de Ch\^{a}teau-Gombert 38, \\ rue Fr\'ed\'eric Joliot-Curie 13388 Marseille cedex 13 FRANCE\\
$^{10}$  ETH Zurich, Institute for Particle Physics and Astrophysics, Wolfgang-Pauli-Stra\ss e 27, CH-8093 Zurich, Switzerland\\
$^{11}$  CRAL, UMR 5574, CNRS, Universit\'e Lyon 1, 9 avenue Charles Andr\'e, 69561 Saint Genis Laval Cedex, France\\
$^{12}$ TAPIR, Walter Burke Institute for Theoretical Physics, Mailcode 350-17 Caltech, Pasadena, CA 91125, USA\\
$^{13}$N\'ucleo de Astronom\'ia, Facultad de Ingenier\'ia y Ciencias, Universidad Diego Portales, Av. Ejercito 441, Santiago, Chile\\
$^{14}$Escuela de Ingenier\'ia Industrial, Facultad de Ingenier\'ia y Ciencias, Universidad Diego Portales, Av. Ejercito 441, Santiago, Chile \\
$^{15}$Aix Marseille Universit\'e, CNRS, LAM - Laboratoire d'Astrophysique de Marseille, UMR 7326, 13388, Marseille, France  \\
}

\date{Accepted XXX. Received YYY; in original form ZZZ}

\pubyear{2018}

\begin{document}
\label{firstpage}
\pagerange{\pageref{firstpage}--\pageref{lastpage}}
\maketitle

\begin{abstract}
In a multi-wavelength study of thermal emission and scattered light images we analyse the dust properties and structure of the debris disc around the A1-type main sequence star 49~Cet. As a basis for this study, we present new scattered light images of the debris disc  known to possess both a high amount of dust and gas. The outer region of the disc is revealed in former coronagraphic H-band and our new Y-band images from the Very Large Telescope SPHERE instrument. 
We use the knowledge of the disc's radial extent inferred from ALMA observations and the grain size distribution found by SED fitting to generate semi-dynamical dust models of the disc.
We compare the models to scattered light and thermal emission data and find that 
a disc with a maximum of the surface density at 110~au and shallow edges 
can describe both thermal emission and scattered light observations. 
This suggests that grains close to the blow-out limit and large grains stem from the same planetesimal population and are mainly influenced by radiation pressure. The influence of inwards transport processes could not be analysed in this study.
\end{abstract}




\section{Introduction}

Circumstellar discs around young stars are natural by-products of star formation. They serve as reservoir for mass accretion when protostars form and afterwards transform to places where planets can form. At the beginning of their evolution, primordial discs are mostly composed of gas and only a minor mass fraction is present in small solid dust particles. The gas plays a major role in controlling the disc dust dynamics \citep{beckwith-et-al-2000}. Due to viscous accretion \citep{lyndenbell-pringle-1974} and photo-evaporation \citep{alexander-et-al-2006} the gas is removed during the discs evolution process and is mostly depleted during the first $\sim$10~Myr \cite[e.g.,][]{zuckerman-et-al-1995, fedele-et-al-2010}. 

After this evolutionary phase, the dynamics of dust particles, being no longer governed or stabilised by the gas, are strongly influenced by stellar radiation pressure and Poynting-Robertson drag. In such environment the lifetime of grains is much shorter than the lifetime of the host star \citep[e.g.,][]{dominik-decin-2003, wyatt-2005}. 
In the last decades observations at infrared wavelengths revealed hundreds of dust dominated discs around stars with a wide range of ages \citep{hughes-et-al-2018}. Considering the limited lifetime, the dust material of these {\sl debris discs} could not be leftover from the primordial stage but rather be comprised of second generation particles continuously replenishing from collisions and evaporation of previously formed larger planetesimals \citep[e.g.,][]{wyatt-2008, krivov-2010}. Besides dust, recent observations revealed gas, mostly CO molecules, in emission in $\sim$20 of these debris systems \citep{hughes-et-al-2018}. Similarly to dust particles the detected gas is also thought to be derived from larger planetesimals and thus having secondary origin \citep{kral-et-al-2017},  though in some very gas-rich, young systems a primordial, residual origin also cannot be ruled out \citep{kospal-et-al-2013}.

49~Cet (HD~9672) is one of the most prominent examples for young, dust-rich gas-bearing debris discs. The A1V-type host star has a stellar luminosity of 16$\Lsun$ and an effective temperature of 9000~K \citep{roberge-et-al-2013}. Its distance is given by the new Gaia data release as $57.0\pm0.3$~pc \citep{GAIA-collaboration-et-al-2016,GAIA-collaboration-et-al-2018b, bailer-jones-et-al-2018}. 49~Cet is found to be a member of the 40~Myr old Argus association \citep{torres-et-al-2008, zuckerman-et-al-2012, zuckerman-2018}. The dust disc around 49~Cet was discovered by the Infrared Astronomical Satellite (IRAS) \citep{neugebauer-et-al-1984, sadakane-nishida-1986} and found to be one of the brightest discs with a fractional luminosity above $10^{-3}$ \citep{jura-et-al-1993}. It was first spatially resolved in the mid-infrared with the MIRLIN instrument of the Keck~II 10~m telescope \citep[e.g.,][]{wahhaj-et-al-2007}. However, in the mid-infrared regime the disc exhibits only a moderate amount of emission, suggesting that the dust needs to have a low temperature. 49~Cet is one of those rare gaseous debris discs, where besides CO gas \citep{zuckerman-et-al-1995} several other gas compounds have been detected either in emission \citep{donaldson-et-al-2013, higuchi-et-al-2017} or in absorption \citep{roberge-et-al-2014}. The origin of this gas is still debated but most current evidences lean towards a second-generation origin \citep[e.g.,][]{hughes-et-al-2017}.

Recently, the results of the numerous studies of the 49~Cet debris disc were summarised in \cite{choquet-et-al-2017}, who presented the first analysis of scattered light images of this disc observed by \textit{Hubble Space Telescope}/NICMOS and \textit{Very Large Telescope}/SPHERE. The authors generated a schematic view of the system and furthermore, they investigated the possibility of existing but yet unseen planets around the 49~Cet host star. In addition to that, \cite{hughes-et-al-2017} presented ALMA images with a spatially resolved surface density distribution of the molecular gas and the dust continuum emission.

Thanks to the wide wavelength coverage of observations of 49~Cet, it is possible to compare its system properties inferred by studies at different wavelengths. 
An interesting question is whether we see similar disc structures for different dust grain populations. By combining near-infrared scattered light emission, tracing smaller grains, and thermal emission in the sub-mm regime, sensitive to large particles, we are able to address this issue.
Furthermore, by combining different kind of observations at multiple wavelengths,
it is possible to constrain system parameters which remained unconstrained in former studies focusing on one wavelength range or type of emission. Due to the developments in observational techniques and instruments this approach became feasible for debris discs only recently, but at the time of this study a handful of analyses addressed this topic already \citep[e.g.,][Olofsson et al., in prep.]{augereau-et-al-2001, ertel-et-al-2011, donaldson-et-al-2013, schueppler-et-al-2015, lebreton-et-al-2016, olofsson-et-al-2016}.
Nevertheless, most resolved discs \citep[146, List of resolved debris discs\footnote{https://www.astro.uni-jena.de/index.php/theory/catalog-of-resolved-debris-disks.html}][]{hughes-et-al-2018} do not have data in both wavelength regimes so that a detailed analysis of the disc extent in scattered light and thermal emission remains a future project. 

Large dust particles are most effectively traced by observations of thermal emission at long wavelengths (e.g., far-infrared or sub-millimetre). Since these grains are not sensitive to radiation pressure forces we assume their position to be close to the planetesimal belt invisible for us. On the other hand, scattered light in the near-infrared is dominated by small particles which are highly affected by radiation pressure and can have highly eccentric orbits as a consequence.
As a result we would expect that by observing small grains we should see a more extended dust disc than by tracing large particles at mm-wavelengths. 
Indeed, for example the debris disc around $\beta$~Pic shows an extent of $\sim$150~au at sub-mm wavelengths while in near-infrared a halo of small particles becomes visible up to 1800~au \citep{ballering-et-al-2016}.
However, for some discs such as Fomalhaut comparable disc radii are found in thermal emission and scattered light \citep{holland-et-al-1998, dent-et-al-2000, kalas-et-al-2005}.
In the case of 49~Cet, we are fortunately able to compare former ALMA studies \citep{hughes-et-al-2017} with observations in scattered light \citep[this work, see also][]{choquet-et-al-2017} to address the question of the disc extent as well.
An overview of the theoretical background for the dust dynamics in debris discs is presented in Appendix~\ref{sec:theory}.

In this paper, we will present new Y-band data taken with VLT/SPHERE and compare them to previous SPHERE H-band data from \cite{choquet-et-al-2017}. The observations and data reduction of the SPHERE data are described in Section~\ref{sec:observations}. We will analyse the radial extent of the disc at different wavelengths in Section~\ref{sec:results}. The modelling of the disc is described in Section~\ref{sec:morphology}.
Furthermore, we will investigate the question of whether it is possible to find a model which can fit the disc parameters inferred from both thermal emission and scattered light images. 
Section~\ref{sec:analysis} shows the analysis of the different models investigated, while the results are discussed in section~\ref{sec:discussion}. A summary is given in section~\ref{sec:summary}.

\section{Observations and data reduction}
\label{sec:observations}

\subsection{Broad band SPHERE data}
We observed the disc around 49~Cet in the programme 198.C-0209(N) on 19$^\text{th}$ November 2016 for 1.8~hours on source with the SPHERE/IRDIS instrument of the VLT \citep{beuzit-et-al-2019,dohlen-et-al-2008}. The observations were carried out using the broad band Y filter with a central wavelength of 1.04$\mum$, a width of 139~nm and the coronagraph N\_ALC\_YJ\_S \citep{martinez-et-al-2009,carbillet-et-al-2011} with a diameter of 185~mas. The observations were performed in pupil tracking mode to allow for angular differential imaging \citep[ADI,][]{marois-et-al-2006}. The total on-sky rotation during our observations was 70$^{\circ}$.

\begin{figure*}
	\includegraphics[width=0.7\textwidth]{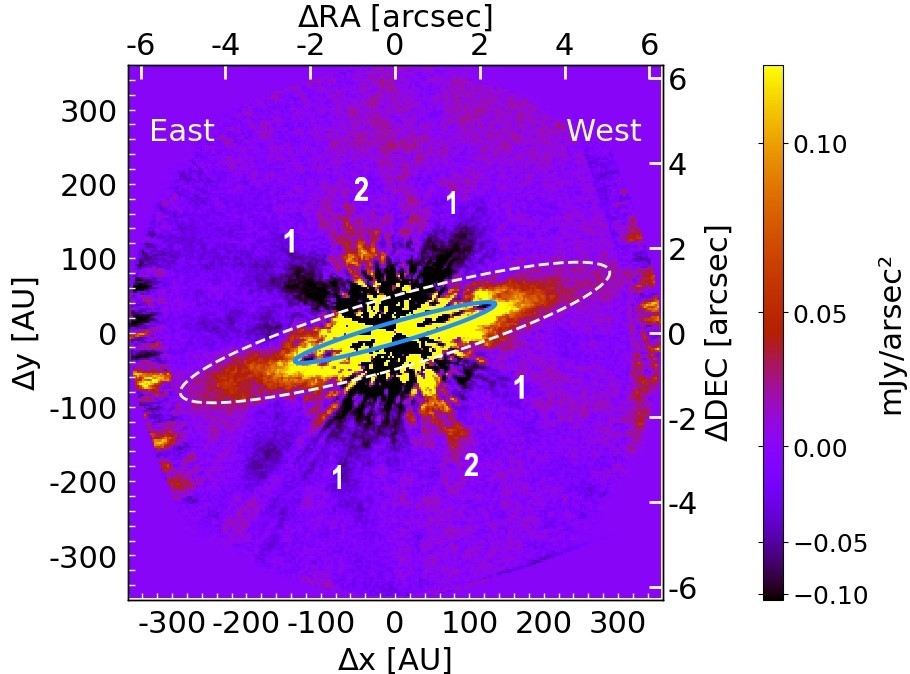}
	\caption{Classical ADI-reduced scattered light Y-band image of 49~Cet. The white dashed line shows the inferred disc extent of 280~au, the blue solid line the location of the surface brightness peak at 140~au. The numbers 1 and 2 indicate the residuals . Further explanations are given in the text.}
	\label{fig:ADI_image}
\end{figure*}

After basic reduction steps (flat fielding, bad-pixel correction, background subtraction, frame registration, frame sorting) we processed the data with a classical ADI reduction technique, which consisted of building a model PSF from the mean of all pupil-stabilised images, which was then subtracted from each frame before de-rotating and stacking the images.

In Fig.~\ref{fig:ADI_image} the result of the ADI-reduction of the 49~Cet debris disc is shown. 
The image was normalised to mJy/arcsec$^2$ in the following way. On the non-coronagraphic image, we measured the flux density encircled within a circle of radius 0.1~arcsec, encompassing the PSF core, wings and diffraction spikes from the spiders. Then this flux density is corrected by the transmission of the neutral density filter used to obtain the non-coronagraphic image, and by the ratio between the Detector Integration Time (DIT) of the coronagraphic and non-coronagraphic images, to obtain a reference conversion value.
To convert the coronagraphic image from ADU to mJy/arcsec$^2$, the coronagraphic image is divided by the reference conversion value, multiplied by the stellar flux density found to be 11.9~Jy at the central wavelength of the Y band and divided by the pixel surface area in arcsec$^2$. The pixel scale of IRDIS is 0.01225 arcsec/pixel \citep{maire-et-al-2016}. 
The ADI reduction algorithm induces self-subtraction of any extended astrophysical signal \citep{milli-et-al-2012}. The flux displayed in Fig.~\ref{fig:ADI_image} did not take this self-subtraction effect into account, which requires a forward-modelling approach and is described in Section~\ref{sec:MODERATO}.

\subsection{Narrow band SPHERE data}

We also observed 49 Cet on 23$^{\text{rd}}$ July 2016, with VLT/SPHERE in the framework of an open time programme 097.C-0747(A). We used pupil-stabilised imaging with the N\_ALC\_YJH\_S coronagraph, having a coronagraphic mask with a diameter of 185~mas. We used the IRDIFS observing mode, which provided IFS observations in the Y-J range and IRDIS data in the H23 dual band. The data reduction was carried out by the SPHERE Data Centre using their pipeline \citep{delorme-et-al-2017}. The obtained reduced master cubes were then utilised as input for high contrast imaging post-processing performed by the Speckle Calibration (SpeCal) package \citep{delorme-et-al-2017, galicher-et-al-2018}. 
\begin{figure}
	\includegraphics[width=\columnwidth]{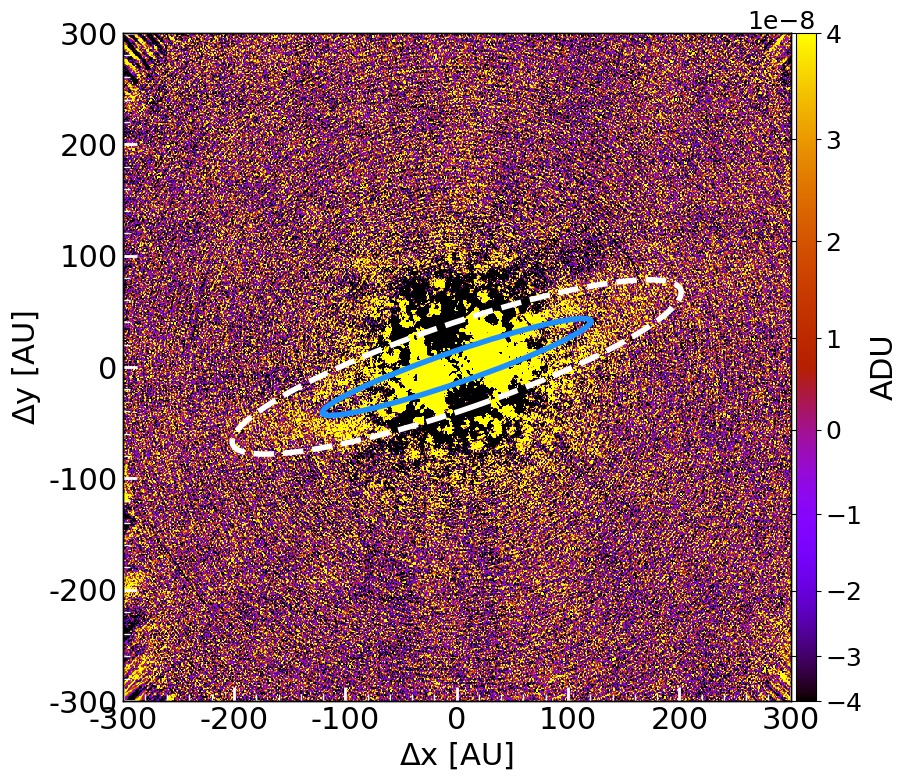}
	\caption{Classical ADI-reduced scattered light H-band image of 49~Cet. The white dashed line shows the inferred disc extension of 200~au, the blue solid line the location of the surface brightness peak at $\sim$140~au.
			}
	\label{fig:ADI_image_NarrowH}
\end{figure}
The reduced image in ADU is depicted in Fig.~\ref{fig:ADI_image_NarrowH}.

\section{Disc geometry and radial profiles}

In our ADI-reduced Y-band image (Fig.~\ref{fig:ADI_image}) we clearly detect the debris disc of 49~Cet between a radial distance of $1.4\arcsec$ (80~au) and $\sim4.9\arcsec$ (280~au). We note that these values are detection limits and thus, might not resemble the true, possibly larger, disc extent. At an angle of $45^\circ$ to the semi-major axis we see dark patterns as residuals from the reduction process (shown as No. 1 in Fig.~\ref{fig:ADI_image}.). In addition to that, there seems to be emission at a $90^\circ$ angle which is identified as diffraction pattern residuals stemming from the star (shown as No. 2 in Fig.~\ref{fig:ADI_image}.). In the reduced H23-band image the disc is detected between 80 and $\sim$200~au where the smaller extent is caused by a weaker detection of the disc. Nevertheless, in contrast to the Y-band data, neither the dark pattern, nor the diffraction spikes are visible in the H23-band. 

In the following study we will focus on the Y-band data due to the stronger detection of the debris disc. Furthermore, we will concentrate on the resolved outer disc component and will not take a possible inner ring into account which was proposed at a location of $\sim10$~au by former studies \citep[e.g.,][]{wahhaj-et-al-2007, chen-et-al-2014, pawellek-et-al-2014}. 

We derived the PA of the disc in Y-band, and found a best value of $106.2^\circ\pm1.0^\circ$. To do so, we de-rotated the disc to align the major axis with the horizontal of the de-rotator. Then we fitted the vertical profiles of the disc between a separation of 45 and 105 pixels with a Gaussian profile. We iterated with the de-rotation angle until the centroids of the Gaussian had a slope of zero \citep[e.g.,][]{lagrange-et-al-2012a, milli-et-al-2014}. Comparing our PA to values from former studies which lie between 93 and 130 degrees (see Tab.~\ref{tab:disc_wavelengths}) we find a good agreement with the Herschel/PACS measurements.
Comparing directly to other SPHERE observations is difficult since the value of 110$^\circ$ inferred by \cite{choquet-et-al-2017} was not fitted but fixed.
\begin{table*}
\caption{Disc parameters inferred in different wavelength ranges.
\label{tab:disc_wavelengths}}
\begin{tabular}{ccccl}
Wavelength [$\mum$]	& Radius [au]	& Inclination [$^\circ$]	& PA [$^\circ$]	& Comments / Instrument \\
\toprule
NIR				& 65 - 250	 & 73			  & 110				& VLT/SPHERE, \cite{choquet-et-al-2017}  \\
NIR             & 80 - 280   &  \ldots & $106\pm1$         & VLT/SPHERE, this work\\
12.5, 17.9		& 30 - 60	 & $60\pm 15$	  & $125 \pm 10$	& Keck/MIRLIN, inner disc component, \cite{wahhaj-et-al-2007} \\
70				& 200		 & $> 44$		  & $105 \pm 1$		& Herschel/PACS, \cite{roberge-et-al-2013}\\
70				& 192		 & $67.4\pm 2.7$  & $109.0 \pm 3.9$	& Herschel/PACS, \cite{moor-et-al-2015a}\\
100				& 196		 & $67.2\pm 2.5$  & $109.4 \pm 4.8$	& Herschel/PACS, \cite{moor-et-al-2015a}\\
160				& 209		 & $56.7\pm 15.5$ & $93.4 \pm 13.9$	& Herschel/PACS, \cite{moor-et-al-2015a}\\
450				& $421\pm16$ & $74\pm13$	  & $130 \pm 10$	& SCUBA2, \cite{holland-et-al-2017}\\
850				& 117		 & $80.6\pm 0.4$  & $109.1 \pm 0.4$ & ALMA, \cite{hughes-et-al-2017}\\
\end{tabular}
\end{table*}

\label{sec:results}

\subsection{Radial profiles of scattered light images}

We extracted the radial profile of the surface brightness using the same method as described in \cite{choquet-et-al-2017} where the authors analyse H-band data of 49~Cet. In this method we produce slices along the semi-major axis with a length of 93~pixels above and under the semi-major axis and a width of 2~pixels. Then we calculate the mean value of the flux density for each slice. With the sizes chosen for the slices we are able to directly compare our Y-band profile to the H-band data of \cite{choquet-et-al-2017}. We estimate the noise level of the images by generating similar slices as for the radial profile itself. Then we rotate these slices by 90$^\circ$ to get the perpendicular direction to the disc. Finally, we calculate the standard deviation of each slice.

\begin{figure}
\includegraphics[width=\columnwidth]{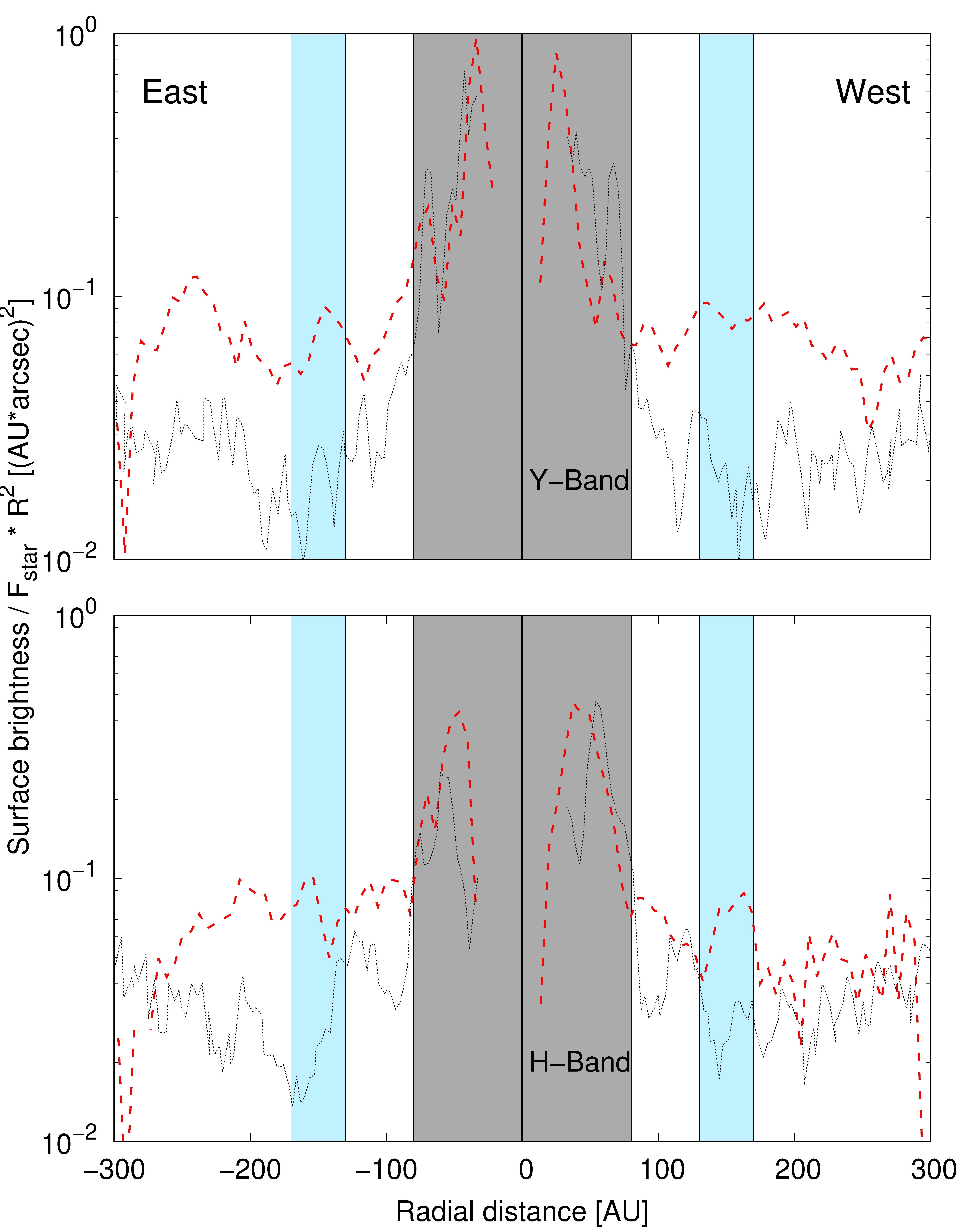}
\caption{Surface brightness as a function of radial distance to the star for the SPHERE observations. The upper panel shows the Y-band data, the lower the H-band data. The red dashed line depicts the radial profile of the disc, the black solid line the $1\sigma$ noise level.Bright blue filled areas in both panels represent the location of symmetric peaks in western and eastern direction. The grey filled area shows the inner region of 80~au where the noise level is comparable to the signal.
}
\label{fig:Radial_Profile}
\end{figure}

The result is shown in Fig.~\ref{fig:Radial_Profile}, where we normalise the surface brightness to the flux density of the host star given as 6.4~Jy for H-band and 11.9~Jy for Y-band.
Our Y-band observations reach a disc signal-to-noise level of 7, which is stronger than for the H-band data due to a longer exposure time. 
The noise that dominates at short separation is the residuals from the subtraction process of the Point Spread Function (PSF). Beyond $\sim3$~arcsec, we are in the background limited regime where the background noise dominates (sky background + readout noise of the detector).
The noise level is in the same order of magnitude as the disc signal within a region of 80~au and thus, we will exclude the inner region from further analyses. Furthermore, the disc is detected within $\sim$280~au based on the noise level as well. 
In case of the H-band data the eastern side of the disc is well detected while the western side remains close to the noise level. In addition, the disc is found to be of a similar extent as suggested by our Y-band image. Analysing the radial profile we find local peaks of the surface brightness in an area between $\sim$130~au and $\sim$170~au in the eastern and western direction in both SPHERE images.

\subsection{Radial profile of thermal emission}

\begin{figure}
\includegraphics[width=0.8\columnwidth, angle=-90]{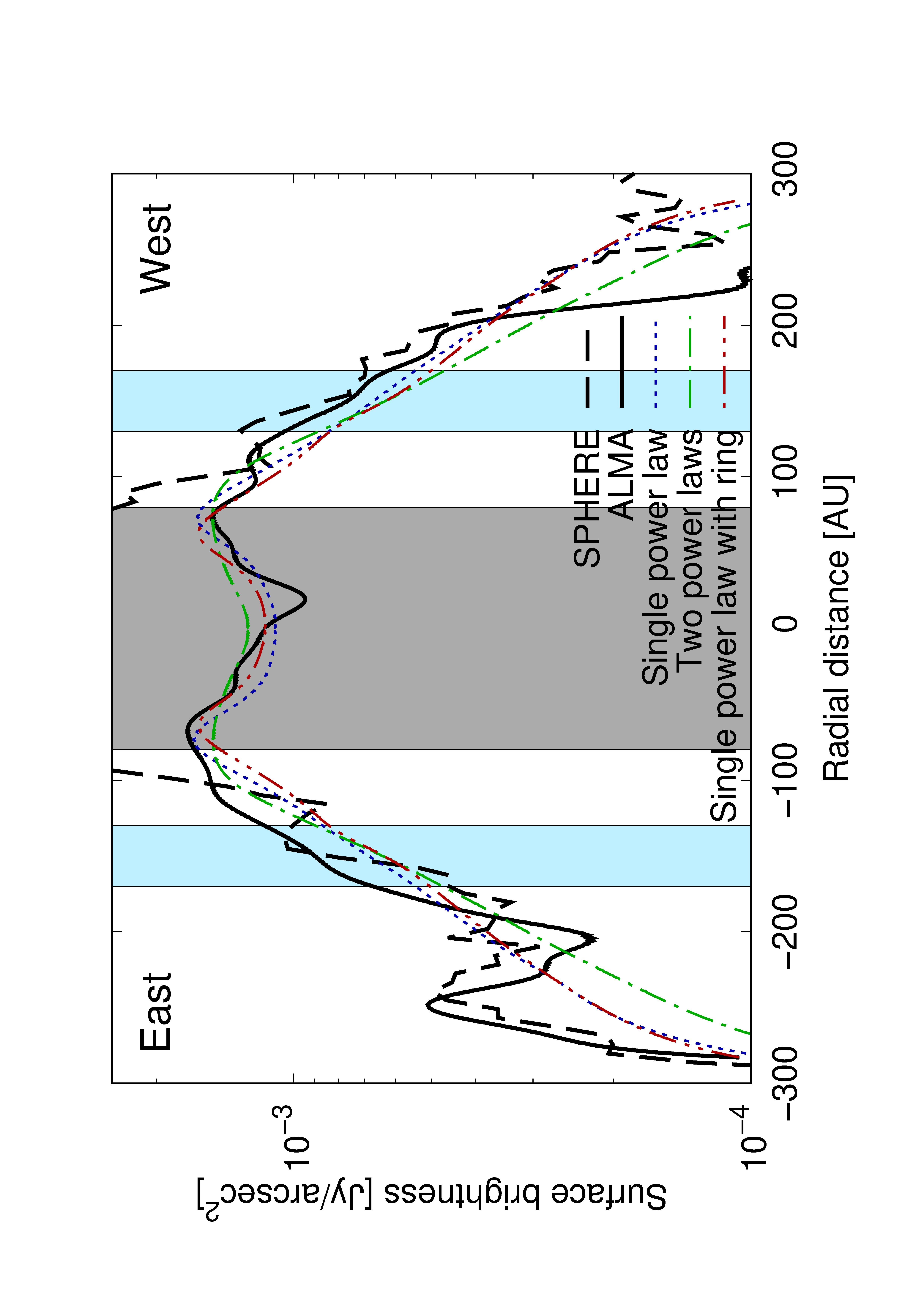}
\caption{Surface brightness as a function of radial distance. The thick black solid line shows the observational data for ALMA, the black dashed line for SPHERE H-band normalised to the ALMA data. The blue dotted line shows the disc with single power law radial profile, the green dash-dotted line a two - power law radial profile and the red dash-double dotted line a single power law including a narrow dust ring. Blue shaded areas
represent symmetric peaks in western and eastern direction at the same position as in Fig.~\ref{fig:Radial_Profile}. 
The grey filled area shows the inner region of 80~au.
}
\label{fig:Radial_Profile_ALMA}
\end{figure}

In order to compare the brightness profile of the near-infrared scattered light with the thermal dust emission in the sub-millimetre, we use ALMA data directly taken from \cite{hughes-et-al-2017} who presented 850$\mum$ continuum images for 49~Cet with a beam size of  $0.47\times0.39$~arcsec. To obtain the radial profile we apply the same method as described for the scattered light observations, but adapt the size of the slices according to the ALMA pixel scale of 9.76562~mas/pixel. 
The resulting radial profile is depicted in Fig.~\ref{fig:Radial_Profile_ALMA}. 
In contrast to the calculated noise level as a function of radial distance for SPHERE data we use the noise level of 0.056~mJy/beam as given in \cite{hughes-et-al-2017} to infer the disc radius.

We find that the eastern wing extends to $\sim$280~au and the western side up to $\sim$250~au which is in agreement with the scattered light data. However, in case of the thermal emission measurements the radial profile shows a clear drop of the surface brightness around $\sim$200~au down to values comparable to the noise level of 0.056~mJy/beam so that we assume to see the real disc extent rather than a radial detection limit.

Comparing our inferred disc extent with values derived by former studies (see Tab.~\ref{tab:disc_wavelengths}) we find comparable disc radii for different Herschel/PACS observations as well as for the SPHERE H-band observation.

\section{Modelling}
\label{sec:morphology}

\subsection{Modelling strategy}

The 49~Cet debris disc system was subject to several modelling projects, either concentrating on the dust \citep[e.g.,][]{wahhaj-et-al-2007, pawellek-et-al-2014, choquet-et-al-2017}, the gas \citep[e.g.,][]{roberge-et-al-2014, miles-et-al-2016} or both components \citep[e.g.,][]{zuckerman-song-2012, roberge-et-al-2013, nhung-et-al-2017, hughes-et-al-2017}. In this study we focus on the dust component.
Our goal is to investigate whether disc models inferred from thermal emission images can explain the scattered light observations obtained in near-infrared.

\begin{figure}
\includegraphics[width=0.8\columnwidth, angle=-90]{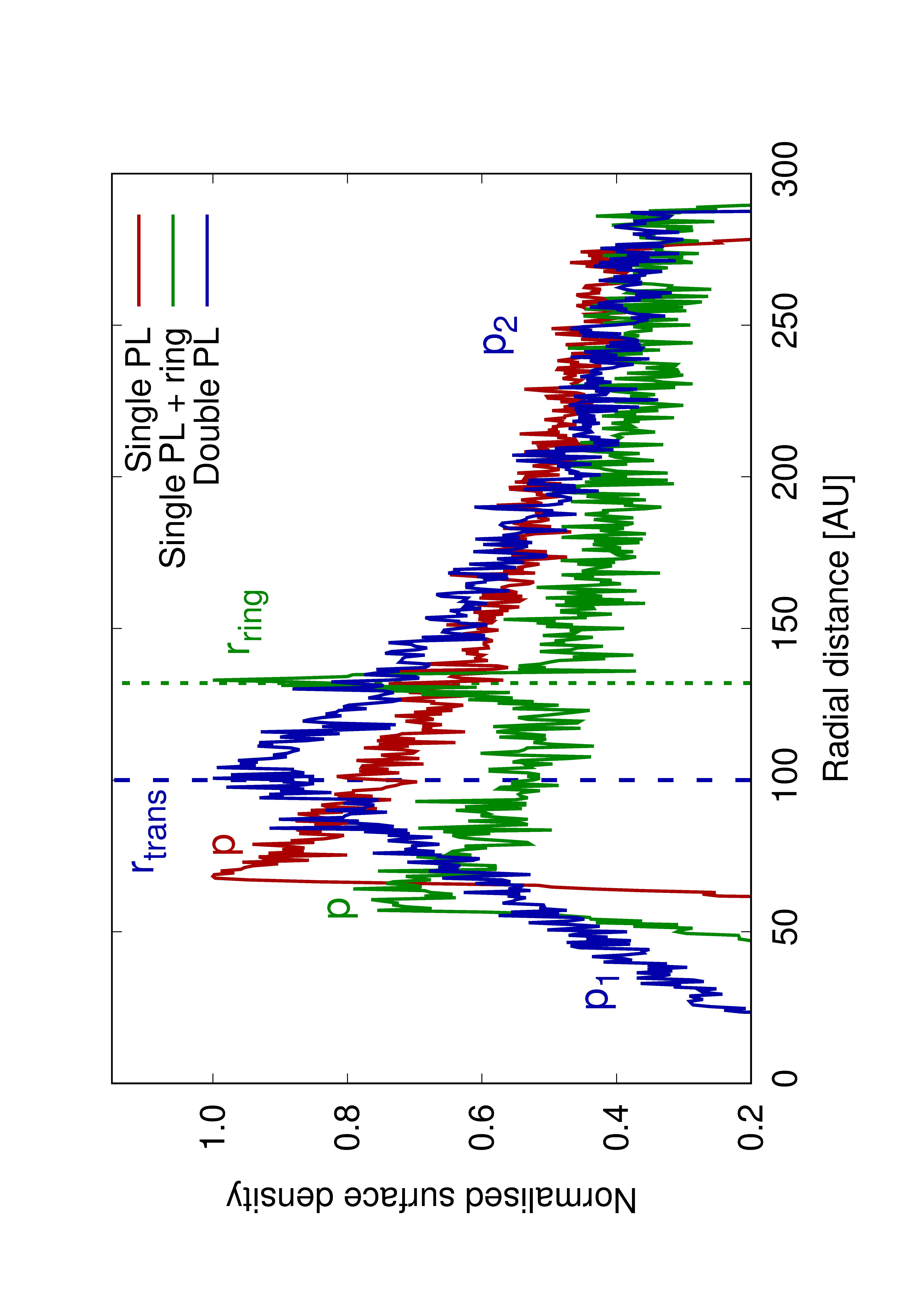}
\caption{Surface density profiles for the parent belts inferred from semi-dynamical models. Red shows the single power law; green the power law including a narrow ring and blue the double power law model. The vertical green dotted line and the blue dashed line give the position of the planetesimal ring and the transition radius, respectively. The different radial distribution indices used for the models are depicted at the corresponding parts of the profiles. The noise is caused by the finite number of planetesimals in the models.
}
\label{fig:SurfaceDensity}
\end{figure}

We start with a fit of the spectral energy distribution (SED) of 49~Cet to get information on the dust grain size parameters (see Section~\ref{sec:SED} for a detailed description).
Then we study the radial distribution of the particles. In a former study \cite{hughes-et-al-2017} investigated three different surface density profiles for the grains visible at sub-mm wavelengths: a single power law, a power law including a narrow dust ring, and a double-power law model.
We assume that large dust particles trace the planetesimals best and thus, use the parameters of the three surface density profiles mentioned before to generate the plantesimal belts for our so-called semi-dynamical disc models.
These models assume that dust grains are released from the planetesimals and that their orbits are altered by stellar radiation pressure. We use the term ``semi-dynamical'' since no time-dependency of moving particles is taken into account.
The actual profiles of the planetesimal belts from which the dust is produced are depicted in Fig.~\ref{fig:SurfaceDensity} while their parameters are specified in Tab.~\ref{tab:radial_parameters}. The noise visible in Fig.~\ref{fig:SurfaceDensity} stems from the number of planetesimals used for the model setting. 
The sizes of the dust grains follow the distribution inferred by the SED modelling. 
From the semi-dynamical models we generate thermal emission and scattered light images.
Details for this procedure are described in Section~\ref{sec:MODERATO}. In a final step the resulting images are compared to the observational images by forward modelling. 
At the time of the ALMA study \citep{hughes-et-al-2017}, the new Gaia distances were not available yet, so we correct the inner and outer disc radii for them.

\subsection{SED modelling}
\label{sec:SED}
The photometric data are collected from the literature and summarised in Table~\ref{tab:photometry}. We use the SONATA code \citep[][]{pawellek-et-al-2014, pawellek-krivov-2015} to fit the SED and apply the same stellar photospheric model used in \cite{pawellek-et-al-2014} to determine the influence of the host star. Here, the stellar temperature, metallicity, and surface gravity are taken into account to generate a synthetic spectrum by interpolating the PHOENIX-GAIA model grid \citep{brott-hauschildt-2005}. The dust composition is assumed to be astronomical silicate \citep{draine-2003} with a bulk density of 3.3~g/cm$^3$.

The SONATA code calculates the temperature and the thermal emission of each dust particle at a given distance to the star, where the number of particles is determined by the dust mass. Then it sums up the emission of all particles to generate the SED.
The flux densities given for wavelengths shorter than 10$\mum$ are not used to fit the dust disc since in this wavelength regime the stellar photosphere rather than the dust dominates the emission. The code assumes a power-law for both the radial and the size distribution of the dust using the surface number density $N(r,s)$:
\begin{equation}
N(r,s) \sim \left(\frac{r}{r_0}\right)^{-p} \times \left(\frac{s}{s_0}\right)^{-q}.
\label{eq:SurfaceNumberDensity}
\end{equation}
Here, $r$ represents the disc radius, $s$ the grain radius, $s_0$ and $r_0$ normalisation factors and $p$ and $q$ the power-law indices for the radial and size distribution. The surface number density is directly connected to the surface density, $\Sigma(r,s)$, by
\begin{equation}
\Sigma(r,s)\text{d}s = \pi s^2 \times N(r,s) \text{d}s.
\end{equation}
Due to the fact that SEDs use photometric data integrated over the whole disc area, they cannot differentiate between different radial profiles. Therefore, we assume the same single power law model given by \cite{hughes-et-al-2017} with $p=1.27$, $r_\text{min} = 70$~au and $r_\text{max} = 298$~au to fit the SED.

We assume the grain sizes to lie between a minimum and a maximum value, $s_\text{min}$ and $s_\text{max}$. 
We fix the maximum grain size to 1~cm, because larger grains do not contribute any more to the SED in the wavelength range observed. Furthermore, we fix the radial parameters and hence, we are left with three free parameters to fit, namely the minimum grain size, $s_\text{min}$, the size distribution index, $q$ and the amount of dust, $M_\text{dust}$, for particles between $s_\text{min}$ and $s_\text{max}$.
We infer the amount of dust by using the bulk density $\varrho$ and the dust volume $V$:
\begin{equation}
M_\text{dust} = \varrho \times V = \varrho \times \frac{4\pi}{3}\int\limits_{s_\text{min}}^{s_\text{max}}2\pi\int\limits_{r_\text{min}}^{r_\text{max}} N(r,s)\,\, r\,dr \,\, s^3\,ds.
\label{eq:DustMass}
\end{equation}

Roughly 2/3 of all discs investigated so far show evidence for an asteroid belt analogue (also called warm component) in addition to an Edgeworth-Kuiper-belt analogue \citep[cold component, e.g.,][]{ballering-et-al-2013, chen-et-al-2014, pawellek-et-al-2014}. We apply the criteria given in \cite{pawellek-et-al-2014} to check for the presence of a warm component for 49~Cet and indeed, our modelling results suggest the existence of an inner dust belt as well. 
Since there is a large degeneracy between the outer and inner disc parameters, we have to make several assumptions for our SED model.
At first we have to identify the location of the warm component which is not spatially resolved in the images. To do so, we fit the warm dust component with a pure blackbody model to infer its blackbody radius. Here we assume the same radial distribution index ($p=1.27$) as for the outer component. The blackbody radius is found to be $\sim8$~au. Now, we estimate the ``true'' disc radius by applying the method presented in \cite{pawellek-krivov-2015} and \cite{pawellek-2017}. Here the authors found a relation between the true disc radius and the blackbody radius in the shape of 
\begin{equation}
    \frac{R_\text{disc}}{R_\text{blackbody}} = A\times\left(\frac{L}{L_\text{sun}}\right)^B.
\end{equation}
For pure astronomical silicate \cite{pawellek-2017} give the parameters as $A = 6.49\pm0.86$ and $B = -0.37\pm0.05$. Using these values and the solar luminosity of 16$\Lsun$ we find an estimated ``true'' location of the warm component of $\sim16$~au.
We assume the same minimum grain size and size distribution index for both the warm and cold component and therefore increase the number of free parameters from three to four by adding the amount of dust of the warm component.
The SONATA code uses the simulated annealing approach \citep{pawellek-et-al-2014} to fit the SED.
Considering the scattered light images, we are not sensitive to the region within 80~au. Thus, we focus on the cold dust component of the debris disc.

\begin{table}
\caption{Parameters for the radial distribution models.
\label{tab:radial_parameters}}
\tabcolsep 3pt
\centering
\begin{tabular}{rr|rr|rr|rr}
\multicolumn{2}{c|}{Parameter}				& \multicolumn{4}{c|}{Single PL}	& \multicolumn{2}{c}{Double PL} 	\\
			&			& \multicolumn{2}{c|}{no ring}			& \multicolumn{2}{c|}{with ring}			&			&		      \\
                &                & H17 & Fit & H17 & Fit & H17 &Fit\\
\toprule
\multicolumn{2}{c|}{$i$ [$^\circ$]	}                & \multicolumn{2}{c|}{79.3}				   &  \multicolumn{2}{c|}{79.3}				   & \multicolumn{2}{c}{79.2}				  \\
\multicolumn{2}{c|}{$PA$ [$^\circ$]	}           & \multicolumn{2}{c|}{108.7}				  & \multicolumn{2}{c|}{108.8}				  & \multicolumn{2}{c}{108.3}				\\
\multicolumn{2}{c|}{$r_\text{min}$ [au]}      &   \multicolumn{2}{c|}{ 70}				  &  \multicolumn{2}{c|}{  58	}			    &  \multicolumn{2}{c}{ 26}				    \\
\multicolumn{2}{c|}{$r_\text{max}$ [au]}	    & \multicolumn{2}{c|}{ 298}				   & \multicolumn{2}{c|}{ 291}				    & \multicolumn{2}{c}{ 302}				   \\
\hline
$p$     & range      & \ldots&0.0-2.0 &\ldots & 0.0-1.5&\multicolumn{2}{c}{\ldots}\\
       & step size & \ldots & 0.2 & \ldots & 0.5 & \multicolumn{2}{c}{\ldots}\\
       & value      & 1.27	&0.6			& 0.75	&		0.5	& \multicolumn{2}{c}{ \ldots}\\
$r_\text{ring}$ [au] & range & \multicolumn{2}{c|}{\ldots} & \ldots & 90-150 & \multicolumn{2}{c}{\ldots}\\
& step size& \multicolumn{2}{c|}{\ldots} & \ldots & 10 & \multicolumn{2}{c}{\ldots}\\
& value& \multicolumn{2}{c|}{ \ldots}			& 109		&	140	&\multicolumn{2}{c}{  \ldots}			\\
$r_\text{trans}$ [au] &range& \multicolumn{2}{c|}{\ldots} & \multicolumn{2}{c|}{\ldots}& \ldots & 50-170\\
&step size& \multicolumn{2}{c|}{\ldots} & \multicolumn{2}{c|}{\ldots}& \ldots & 20\\
& value	  & \multicolumn{2}{c|}{ \ldots}			  & \multicolumn{2}{c|}{ \ldots}			     & 92		& 110		\\
$p_1$&range& \multicolumn{2}{c|}{\ldots} & \multicolumn{2}{c|}{\ldots}& \ldots & -3.0-0.0\\
&step size& \multicolumn{2}{c|}{\ldots} & \multicolumn{2}{c|}{\ldots}& \ldots & 0.5\\
& value & \multicolumn{2}{c|}{ \ldots}			& \multicolumn{2}{c|}{ \ldots}			& -2.7		& -1.0		\\
$p_2$&range		& \multicolumn{2}{c|}{ \ldots}			& \multicolumn{2}{c|}{ \ldots}			& \ldots	& 0.0-2.0			\\
&step size		& \multicolumn{2}{c|}{ \ldots}			& \multicolumn{2}{c|}{ \ldots}			& \ldots	& 0.5		\\
&	value				& \multicolumn{2}{c|}{ \ldots}			& \multicolumn{2}{c|}{ \ldots}			& 1.50	& 1.0			\\
$\chi^2_\text{red}$ &thermal            & 1.40 & 1.18 & 1.30 & 1.16 & 1.32 & 1.15 \\
&scattered & 2.38 & 2.11 & 2.37 & 2.16 & 2.37 & 1.99 \\
\end{tabular}

\noindent
{\em Notes:}
The abbreviation PL stands for power law. The parameter values of the first table part were taken from \cite{hughes-et-al-2017}, corrected for the new distance of 57.1~pc, and used for the modelling of this work.
In the second part of the table the values in column H17 stem from \cite{hughes-et-al-2017} and are corrected for the new distance of 57.1~pc. 
Parameters in the column Fit are the best fit results of this work. 

\end{table}

\begin{table}

\caption{Continuum flux density.
\label{tab:photometry}}
\tabcolsep 2pt
\begin{tabular}{rrclcc}

Wavelength      & \multicolumn{3}{c}{Flux density}       & Instrument & Reference				\\
$[\mum]$        & \multicolumn{3}{c}{[mJy]}              &            &           			\\
\toprule
0.44            & 22260  & $\pm$ & 350    & TYCHO                 & 1			\\    
0.55            & 21600  & $\pm$ & 200    & TYCHO                 & 1			\\    
1.24            & 10180  & $\pm$ & 190    & 2MASS                 & 2			\\   
1.65            & 6300   & $\pm$ & 130    & 2MASS                 & 2			\\  
2.16            & 4373   & $\pm$ &  81    & 2MASS                 & 2			\\
3.35            & 1990   & $\pm$ & 120    & WISE                  & 3			\\
4.60            & 1294   & $\pm$ &  49    & WISE                  & 3			\\
5.86            & 690    & $\pm$ &  70    & IRS                   & 4			\\
7.07            & 480    & $\pm$ &  50    & IRS                   & 4			\\
8.97            & 320    & $\pm$ &  30    & IRS                   & 4			\\
9.0				& 366	 & $\pm$ &  13.5  & AKARI/IRC			  & 5			\\
10.8	        & 250    & $\pm$ &  50    & Keck/MIRLIN           & 6			\\
11.40           & 210    & $\pm$ &  20    & IRS                   & 4			\\ 
11.56           & 211    & $\pm$ &  21    & WISE                  & 3			\\
12.0            & 330    & $\pm$ &  66    & IRAS		          & 6			\\
12.50           & 200    & $\pm$ &  26    & Keck/MIRLIN           & 8			\\   
13.90           & 180    & $\pm$ &  20    & IRS                   & 4			\\  
17.90           & 186    & $\pm$ &  25    & Keck/MIRLIN           & 8			\\
18.0			& 199	 & $\pm$ &  16	  & AKARI/IRC			  & 5			\\
22.09           & 238    & $\pm$ &  24    & WISE                  & 3			\\ 
24.00           & 259    & $\pm$ &  10    & Spitzer/MIPS          & 9			\\
25.0            & 380    & $\pm$ &  76	  & IRAS		          & 7			\\
60.0	        & 2000   & $\pm$ & 400	  & IRAS		          & 7			\\
63.19           & 2090   & $\pm$ & 350    & Herschel/PACS Spec    & 9			\\    
70.00           & 2163   & $\pm$ & 151    & Herschel/PACS         & 10			\\
71.42			& 1749	 & $\pm$ & 123	  & Spitzer/MIPS		  & 10			\\  
72.84           & 1950   & $\pm$ & 320    & Herschel/PACS Spec    & 9			\\
78.74           & 1900   & $\pm$ & 310    & Herschel/PACS Spec    & 9			\\
90.0			& 1776	 & $\pm$ & 295	  & AKARI/FIS			  & 5			\\
90.16           & 1880   & $\pm$ & 320    & Herschel/PACS Spec    & 9			\\
100.0           & 1910   & $\pm$ & 380	  & IRAS		          & 7			\\
100.0			& 1919	 & $\pm$ & 134	  & Herschel/PACS		  & 10			\\
145.54          & 1160   & $\pm$ & 180    & Herschel/PACS Spec    & 9			\\    
150	            & 750    & $\pm$ & 500    & ISO                   & 4,11		\\
157.68          & 980    & $\pm$ & 130    & Herschel/PACS Spec    & 9			\\ 
160.00          & 1066   & $\pm$ &  75    & Herschel/PACS         & 10			\\
170             & 1100   & $\pm$ & 500    & ISO                   & 4,11		\\
250.00          & 372    & $\pm$ &  27    & Herschel/SPIRE        & 9			\\
350.00          & 180    & $\pm$ &  14    & Herschel/SPIRE        & 9			\\
450.0           & 125    & $\pm$ &  18    & JCMT/SCUBA-2          & 12			\\
500.00          & 86     & $\pm$ &   9    & Herschel/SPIRE        & 9			\\
850.0           & 17     & $\pm$ &   3    & ALMA                  & 4			\\
850.00          & 13.5   & $\pm$ & 1.5    & JCMT/SCUBA-2          & 12			\\
1200.00         & 12.7   & $\pm$ & 2.8    & IRAM                  & 13			\\
1300.0          & 2.1    & $\pm$ & 0.7    & CARMA                 & 4			\\
1330			& 5.5	 & $\pm$ & 0.7	  & ALMA/ACA			  & 14			\\
9000.0          & 0.0251 & $\pm$ & 0.0055 & VLA                   & 15			\\
\bottomrule

\end{tabular}

\noindent
{\em References:}
[1] - \cite{hog-et-al-2000}; 
[2] - 2MASS All-Sky Catalog of Point Sources; 
[3] - \cite{wright-et-al-2010}; 
[4] - \cite{hughes-et-al-2017};
[5] - AKARI All-Sky Survey Bright Source Catalog
[6] - \cite{jayawardhana-et-al-2001};
[7] - IRAS Faint Source Catalog;
[8] - \cite{wahhaj-et-al-2007};
[9] - \cite{roberge-et-al-2013};
[10] - \cite{moor-et-al-2015a}
[11] - ISO;
[12] - \cite{holland-et-al-2017};
[13] - \cite{bockelee-morvan-et-al-1994};
[14] - Moór et al. (in prep)
[15] - \cite{macgregor-et-al-2016};

\end{table}
 
\begin{figure}
\includegraphics[width=0.8\columnwidth, angle = -90]{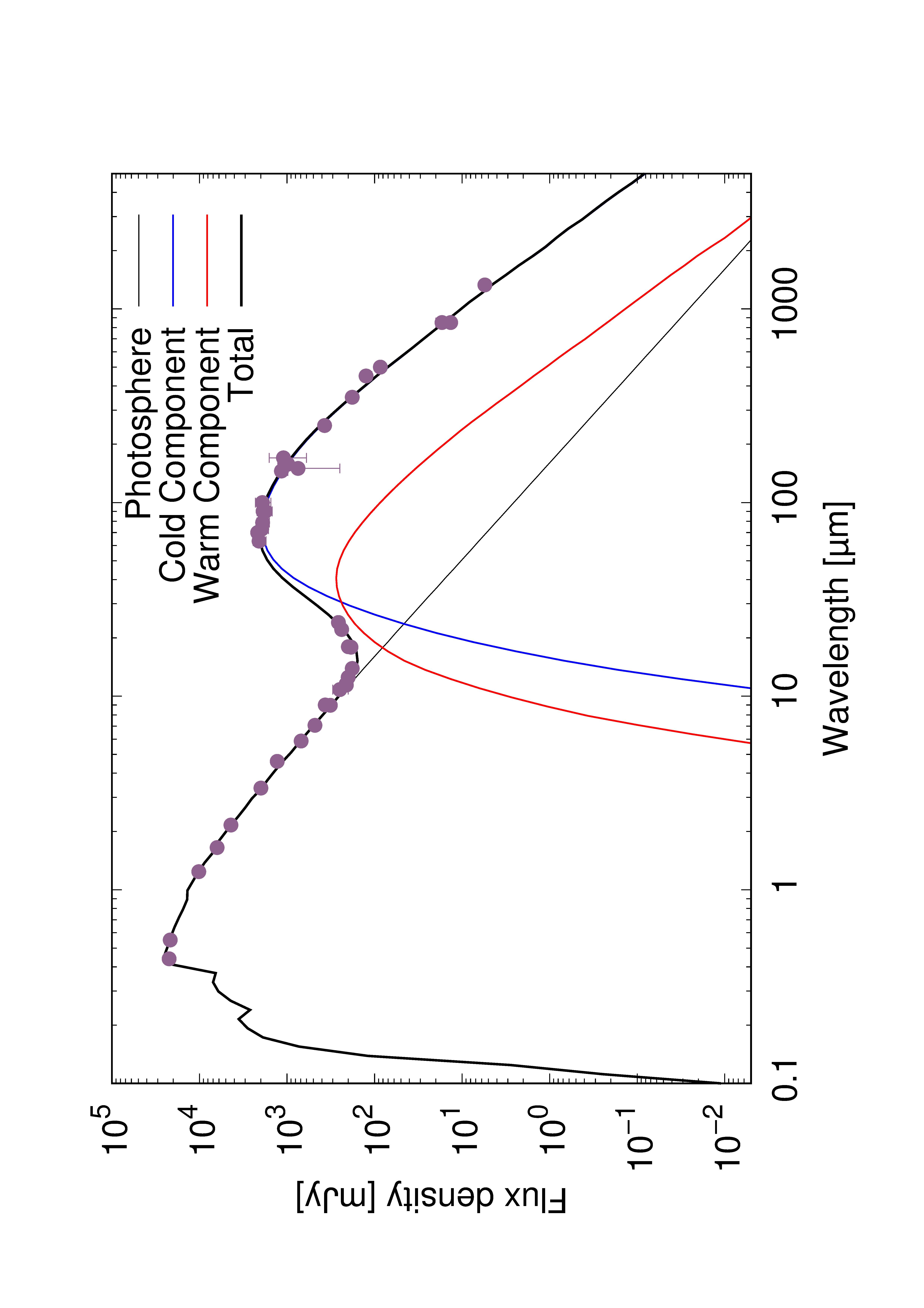}
\caption{SED of 49~Cet using pure astronomical silicate with a bulk density of 3.3g/cm$^3$ for the outer and inner component.
The thin black solid line represents the stellar photosphere, the red solid line shows the warm inner ring, the blue lines the outer belt. 
}
\label{fig:SED}
\end{figure}

In the best fitting SED model the minimum grain size is $s_\text{min} = 5.14\pm 0.76\mum$ and the size distribution index $q = 3.77\pm 0.05$. 
These parameters are applied to all semi-dynamical models with the different radial profiles analysed in this work. We checked the validity of our assumption by using the double power law as radial profile for the SED (Tab.~\ref{tab:radial_parameters}) and found the grain size parameters to be close to the values of the single power law model lying within their given confidence intervals.
The resulting SED is shown in Figure~\ref{fig:SED} while the grain size parameters are given in Table~\ref{tab:grain_parameters}.
The amount of dust, $M_\text{dust}$ , in debris discs depends on the SED model settings, i.e., the grain size and radial distribution and is therefore difficult to compare for different studies. Hence, we focus on the fractional luminosity which is only determined by photometric data points. It is given as $f_\text{d}=8.8\times10^{-4}$.

\begin{table}
\caption{Parameters for the grain size distribution assuming a single power law radial profile.
\label{tab:grain_parameters}}
\tabcolsep 4pt
\centering
\begin{tabular}{r|rl}
Parameter				& Best fit values	& Comment		           \\
\toprule
$s_\text{min}$ [$\mum$] & $5.14\pm0.76$     & \ldots                   \\
$s_\text{max}$ [$\mum$] & 10000 		    & thermal emission, fixed  \\
						& 15	            & scattered light, fixed   \\
$q$	                    & $3.77\pm 0.05$    & \ldots                   \\
$f_\text{d}$			& $8.8\times10^{-4}$& \ldots                   \\
$T_\text{dust}$	[K]		& $65\pm3$	        & \ldots                   \\
$T_\text{warm}$ [K]     & $125\pm2$         & \ldots                   \\
$\chi^2_\text{red}$	    & 1.03              & \ldots                   \\
\end{tabular}

\noindent
{\em Notes:}
The blow out limit is given for pure astronomical silicate as $2.9\mum$.
The different maximum grain sizes are explained in Section~\ref{sec:MODERATO}.
For the dust temperature of both components we take the wavelength of the SED peak and calculate the temperature by using Wien's displacement law.

\end{table}

The error bars give the uncertainties of the fit parameters and are inferred in the following way. We start from the position of the minimum $\chi^2$ in the parameter space. New parameter values are generated and the resulting $\chi^2$ of the model is compared to its minimum value. There is a probability, that a new minimum can be found in the direction of the new parameter values. If this probability reaches a critical value then the fit parameters are saved. In the end, it is counted how often the code reaches a certain parameter value. The resulting distribution in the parameter space represents an estimate for the probability distribution of the parameters and thus, allows us to calculate the confidence levels for the parameters assuming that the values follow a normal distribution \citep[simulated annealing, e.g.,][]{pawellek-2017}.

\subsection{Generating images}
\label{sec:MODERATO}
 
After fitting the SED we generate the semi-dynamical disc models using the MODERATO-code \citep[Olofsson et al., in prep., for similar approaches see][]{wyatt-et-al-1999, Lee-Chiang-2016}. 
The code assumes surface density profiles for the parent belt 
where each planetesimal releases grains of different sizes following a power law (see Eq.~\ref{eq:SurfaceNumberDensity}). Then the orbits of the individual dust particles altered by stellar radiation pressure are calculated
and from their position the scattered light and thermal emission are inferred. 
We correct the possible overabundance of highly eccentric small particles in the way described by \cite{strubbe-chiang-2006} and \cite{thebault-wu-2008}.

The radial and grain size distribution assumed for the models are given in Tabs.~\ref{tab:radial_parameters} and \ref{tab:grain_parameters}.
For each radial profile a small grid of parameter values is assumed to find the best fitting model. Due to the high calculation time for one run, Monte-Carlo fitting is not applicable. 
In all cases we fix the inclination, PA, and the minimum and maximum disc radius to the values inferred by \cite{hughes-et-al-2017} to generate the thermal emission maps. Considering the scattered light maps we use the PA of 106.2 inferred in this work.

In the first scattered light study of 49~Cet, \cite{choquet-et-al-2017} use the anisotropic scattering approach by \cite{henyey-greenstein-1941} where a single scattering asymmetry factor,$|g|$, is assumed for the whole material without a differentiation between grain sizes.
In scattered light studies of debris discs, this approach is often used to get a general idea of the scattering properties of the disc material \citep[e.g.,][]{schneider-et-al-2006, thalmann-et-al-2013, millar-blanchaer-et-al-2015, olofsson-et-al-2016, engler-et-al-2019}. 
Since we generate a physical model with a distribution of particles inferred by SED fitting, we use Mie theory instead to stay consistent and assume pure astronomical silicate \citep{draine-2003} as dust composition similar to the SED modelling. However, we additionally use the Henyey-Greenstein approach to compare our results to previous studies.

We are aware that using Mie-theory to calculate the optical properties for scattered light models leads to an overestimation of forward scattering, especially for larger particles ($\geq20\mum$) since the grains are assumed to be compact spheres 
\citep[e.g.,][]{schuerman-et-al-1981, bohren-huffman-1983, weiss-wrana-1983, mugnai-et-al-1986, mcguire-hapke-1995}. 

To avoid this effect we assess the maximum grain size still contributing to the scattered light.
This is done by estimating the width of the forward scattering peak.
The minimum scattering angle achievable for a disc is given by the disc inclination:
\begin{equation}
\alpha = 90^\circ - i,
\label{eq:ScatteringAngle}
\end{equation}
which is $\approx10^\circ$ in the case of 49~Cet.
On the other hand, analogue to a telescope, a dust particle of radius $s$ can ``detect'' the incoming light (Babinet's principle) with the geometric area $\pi s^2$ and therefore, the Rayleigh criterion provides the minimum angle observable for the grain at a certain wavelength $\lambda$:
\begin{equation}
\alpha = 1.22\times \frac{\lambda}{2\times s} = 70^\circ \times \frac{\lambda}{2\times s}.
\label{eq:GrainAngle}
\end{equation}
Equalising the angles of Eqs.~\ref{eq:ScatteringAngle} and \ref{eq:GrainAngle}, it is possible to infer a grain radius for which the scattered light observations are still sensitive. 
We get an estimate for the maximum particle size of the 49~Cet debris disc as:
\begin{equation}
s  \approx \frac{\lambda\times 70^\circ}{2\times(90^\circ - i)} = \frac{1.04\mum\times70^\circ}{2\times10^\circ} = 3.5\mum.
\end{equation}
This shows, that the scattered light contribution of particles larger than $3.5\mum$ is minor, which is in agreement with \cite{zubko-2013}. In this study, the authors analysed the contribution of large particles in cometary dust by using the Discrete Dipole Approximation method (DDA). They found that for a dust composition with moderate or high absorption, such as carbonaceous material, grains with a size parameter $x = 2\pi \times s/\lambda > 15$ are not needed to model the scattered light observations. However, adding large grains to the model did not change the results significantly, although the computation time was much higher. 

With SED fitting we found a minimum grain size of $\sim5\mum$ which is not contradictory to the above mentioned estimate. It just states that the larger particles are not as effectively traced as smaller grains.
Due to possible estimation uncertainties we set the maximum particle size for scattered light observations to a value of $15\mum$.

The semi-dynamical models are compared to scattered light observations by 
subtracting the convolved disc model from our pupil-stabilised cube of frames. Then we perform ADI on this cube to get the final image. In case of thermal emission, the models are convolved with the assumed ALMA beam and then subtracted from the observational image. 
We use a scaling factor to adapt the flux density of the model to our observations for scattered light and thermal emission. This is inevitable since calculating the flux density of all dust particles necessary to equal the measured emission exceeds the computational resources available.
Due to the different subtraction processes we note that the scaling factors are different for scattered light and thermal emission.
To estimate the quality of the model matching the observations we use $\chi^2$-minimisation assuming that an ideal residual image should expose no emission in any pixel.
The $\chi^2$-parameter is then computed for each pixel by
\begin{equation}
\chi^2 =  \sum\limits_{i=1}^{N_\text{pixel}} { \left(\frac{F_\text{i}}{F_\text{noise}}\right)^2 }
\label{eq:chi2}
\end{equation}
In case of the scattered light data, the noise (or error) is estimated in the following way. We use the noise level inferred from the observational image (see Fig.~\ref{fig:Radial_Profile}). Then we rotate this profile around the centre of the image to generate a noise map.
In case of the thermal emission image we use the noise level given as $F_\text{noise} = 0.056$~mJy/beam \citep{hughes-et-al-2017}. 
The number of pixels for the thermal emission maps is $1024\times1024$ and for the scattered light images $255\times255$.

\section{Analysis}
\label{sec:analysis}
\subsection{Thermal emission}

\begin{figure*}
	\includegraphics[width=\textwidth]{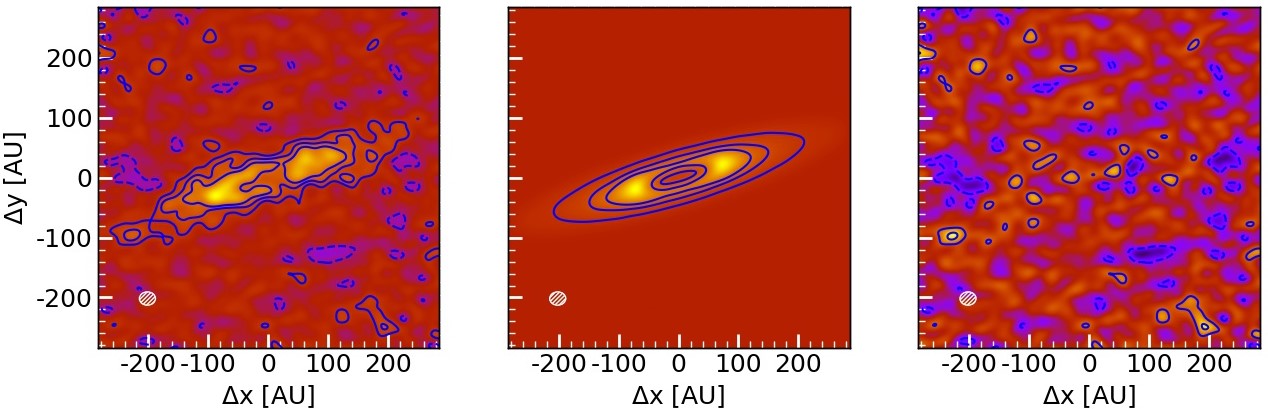}
\caption{Modelling results of 49~Cet. Left: ALMA image at 850$\mum$. Middle: Best fitting model using a single power law radial distribution with $p=0.6$ generated with MODERATO. A grain size distribution with $s_\text{min}=5.14\mum$, $s_\text{max}=1.00$cm and $q=3.77$ was used to generate the disc. Right: Subtraction of the model disc from the observational data. The contour levels show 2, 4 and 6 times the $\sigma$ level given as 0.056~mJy/beam in Hughes et al. (2017). The dotted contours give the negative scale, the white dashed area the beam size. The data are normalised to their maximum. The value of $\chi^2_\text{red}$ is 1.18.}
\label{fig:ModelResults_ALMA_Model1}
\end{figure*}

\begin{figure*}
	\includegraphics[width=\textwidth]{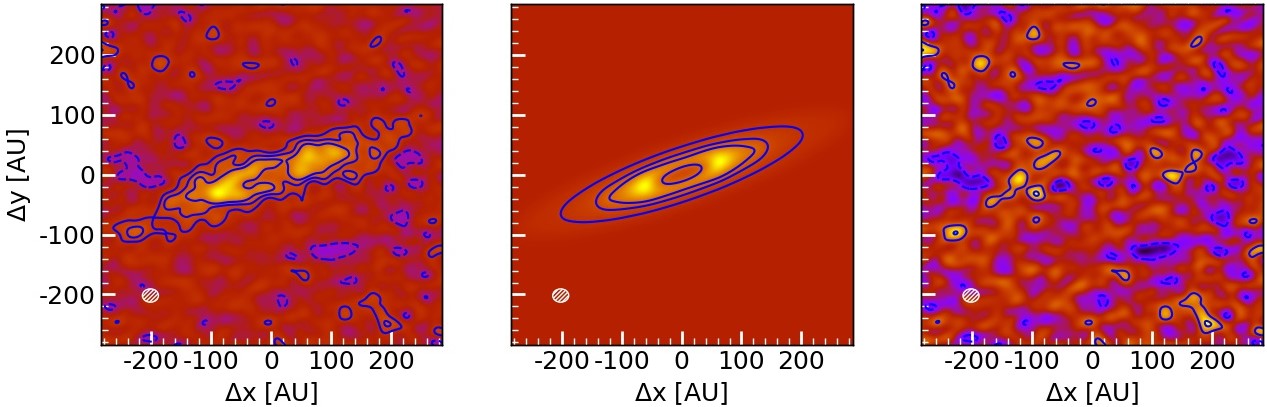}
\caption{Same as Fig.~\ref{fig:ModelResults_ALMA_Model1}, but now using the single power law radial distribution with $p = 0.5$ including an additional narrow ring at $r_\text{ring} = $ 140~au. For this model $\chi^2_\text{red} = 1.16$.}
\label{fig:ModelResults_ALMA_Model2}
\end{figure*}

\begin{figure*}
	\includegraphics[width=\textwidth]{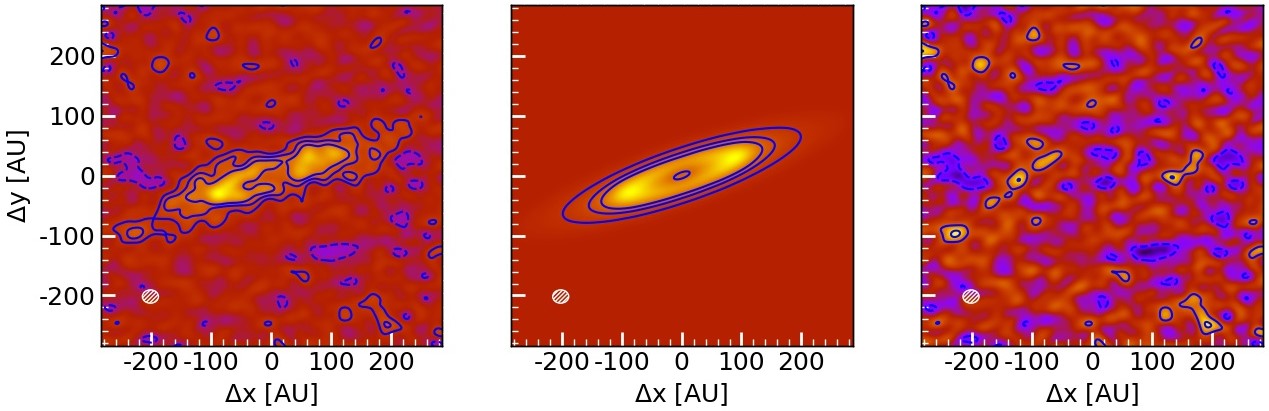}
\caption{Same as Fig.~\ref{fig:ModelResults_ALMA_Model1}, but now using the double power law radial profile with a transition radius of $r_\text{trans} = 130$~au and radial distribution indices of $p_1 = -1.0$ and $p_2 = 1.0$.  The best fit $\chi^2_\text{red} $ is 1.15.}
\label{fig:ModelResults_ALMA_Model3}
\end{figure*}

The thermal emission image observed by \cite{hughes-et-al-2017} with ALMA at 850$\mum$ is compared to the semi-dynamical models based on the three different radial profiles given in Tab.~\ref{tab:radial_parameters} and illustrated in Fig.~\ref{fig:SurfaceDensity}. The best fit results are shown in Figs.~\ref{fig:ModelResults_ALMA_Model1}-\ref{fig:ModelResults_ALMA_Model3}.


\subsubsection{Single power law distribution}

At first a single power law model is assumed as radial distribution. Using the parameter values from \cite{hughes-et-al-2017}, where the disc is confined between 70 and 298~au and possesses a surface density power law exponent of 1.27, we get a reduced $\chi^2$ of $\chi^2_\text{red} = 1.4$.

The grain parameters derived by SED-fitting are applied to our model, where the minimum grain size is 5.14$\mum$, the maximum grain size 1.0~cm and the distribution index $q = 3.77$.  

Next, we analyse a small parameter grid. Here, we fix the minimum and maximum disc radius as well as the inclination and PA, while only one parameter, $p$, is varied between 0.0 and 2.0 in steps of 0.2 .
The result of this fit is shown in Fig.~\ref{fig:ModelResults_ALMA_Model1}, where $p = 0.6$ is the best fit value. We could achieve $\chi^2_\text{red} = 1.18$.
After subtracting the model from the observed data, there are some $2\sigma$ residual structures left throughout the disc. However, they do not seem to be of systematic origin.

\subsubsection{Single power law distribution with a narrow ring}

In a second approach a narrow dust ring is added to the single power law of the radial distribution. The parameter values can be found in Tab.~\ref{tab:radial_parameters}. Here, $N(r,s)$ of the extended disc is calculated at the location of the ring. Then the surface number density of the ring is set to a value twice as high which is in agreement with the parameters given by \cite{hughes-et-al-2017}.
In the final step, the grains of both model parts (disc + ring) are summed up to derive the thermal emission map. 

The parameters given by \cite{hughes-et-al-2017} lead to $\chi^2_\text{red} = 1.30$ which is an improvement to their pure single power law model. 
The ring is confined between 108 and 110~au and possesses the same grain size parameter values as the extended disc between 58 and 291~au.

Fitting a grid of models, we vary two parameters: the distribution index $p$ between 0.0 and 1.5 in steps of 0.5 and the position of the included ring ($r_\text{ring}$) between 90 and 150~au in steps of 10~au. 
The results are shown in Fig.~\ref{fig:ModelResults_ALMA_Model2} where we found a best fitting model with $p=0.5$ and a ring position of $r_\text{ring} = 140$~au. The parameter $\chi^2_\text{red}$ is 1.16 and thus, slightly smaller than for the pure single power law model.  
Analysing the residual image (right panel) there are less residuals found in the inner part of the disc in contrast to the single power law model while in the outer region some larger residuals are left.

\subsubsection{Double power law distribution}

The third approach contains a double power law distribution as radial profile. 
To implement this setting, a similar approach is used as for the former model. We calculate $N(r,s)$ at the position of the transition radius, $r_\text{trans}$. At this position, one power law changes into the other. We demand that $N(r,s)$ has to be the same value for both parts of the radial profile to ensure a continuous surface density profile.

\cite{hughes-et-al-2017} inferred a transition radius between both profile parts at 92~au while the inner disc region is given with $p_1=-2.7$ and the outer fixed to $p_2 = 1.5$. Using these values we get a $\chi^2_\text{red}$ of 1.32. 

Now we vary the transition radius between 50 and 170~au in steps of 20~au and the radial distribution indices between -3.0 and 0.0  for the inner disc region and 0.0 and 2.0 for the outer one in steps of 0.5 each, leading to three free parameters. The best fitting model yields $r_\text{trans}=110$~au, $p_1=-1.0$ and $p_2=1.0$ and leads to $\chi^2_\text{red}=1.15$. The result is depicted in  Fig.~\ref{fig:ModelResults_ALMA_Model3}. 
Comparable to the single power law model including a narrow ring the inner disc region shows no systematic residuals while in the outer part some $2\sigma$ remnants are left. 

\subsubsection{Comparing the models}

By comparing the different radial profiles we find that all models achieve a similar $\chi^2_\text{red}$ value between 1.15 and 1.18. Thus, a conclusion of which model gives the best result is hardly possible. 
Therefore, we apply the Bayesian information criterion ($BIC$) being defined as
\begin{equation}
BIC = \chi^2 + J\times\log_e(N),
\end{equation}
where, $J$ gives the number of free parameters and $N$ the number of data points. The resulting values for each model are given in Tab.~\ref{tab:ThermalEmission_parameters}. 

\begin{table}
\caption{Comparison of thermal emission models
\label{tab:ThermalEmission_parameters}}
\tabcolsep 3pt
\centering
\begin{tabular}{rlrlcl}
No. &Model				                 &$J$	   & Free parameters        & $\chi^2$ [$\times10^6$]	              & $ BIC$ [$\times10^6$]		\\
\toprule
1&SPL             & 2	& scaling, $p$                           & $1.24$ & 1.24\\
2&SPL + ring  &3		& scaling, $p$, $r_\text{ring}$             & $1.22$ & 1.22\\
3&DPL           &4     & scaling, $p_1$, $p_2$, $r_\text{trans}$      & $1.21$ &  1.21\\
\end{tabular}

\noindent
{\em Notes:}
SPL is the single power law model, SPL + ring the single power law including a narrow ring and DPL the double power law model.
The number of data points assumed is the number of pixels: $N = 1024 \times 1024$. 
\end{table}

Following the classification given in \cite{kass-raftery-1995} we compare each model by calculating the parameter  $B = 2\times \log_{e}(\Delta BIC)$, where $\Delta BIC$ represents the difference of $BIC$ between the models. If $B$ lies between 0 and 2 the more complex model is not significantly better than the simple model, by values between 2 and 6 it is possible that the more complex model is better, and between 6 and 10 the more complex model is more likely. For values larger than 10 the probability of the more complex model is much higher than for the simple one. 

We now compare the models 2 and 3 (see Tab.~\ref{tab:ThermalEmission_parameters}) with the single power law model. 
Model 2 reaches a value of $B=19.9$, while model 3 gives $B=20.1$. This indicates that the single power law model including a ring and the double power law model have a much higher probability than the single power law model. 
In the final step we compare model 2 with model 3 and find $B = 18.5$. Thus, although the difference in $\chi^2_\text{red}$ is only minor, the information criterion clearly states that the double power law model is preferable to the power law model including a narrow ring. 

Our fitting results deliver slightly different values than provided by the model of \cite{hughes-et-al-2017}. The main difference here is that in the aforementioned study the applied surface density models for the dust grains do not consider a grain size distribution or any forces altering the particle orbits. We assume a dust orbital model including radiation pressure and a particle size distribution to generate thermal emission maps. 
Nevertheless, both approaches show that a single power law distribution might not be the appropriate way to simulate the debris disc of 49~Cet.

\subsection{Scattered light}

\begin{figure*}
\includegraphics[width=\textwidth]{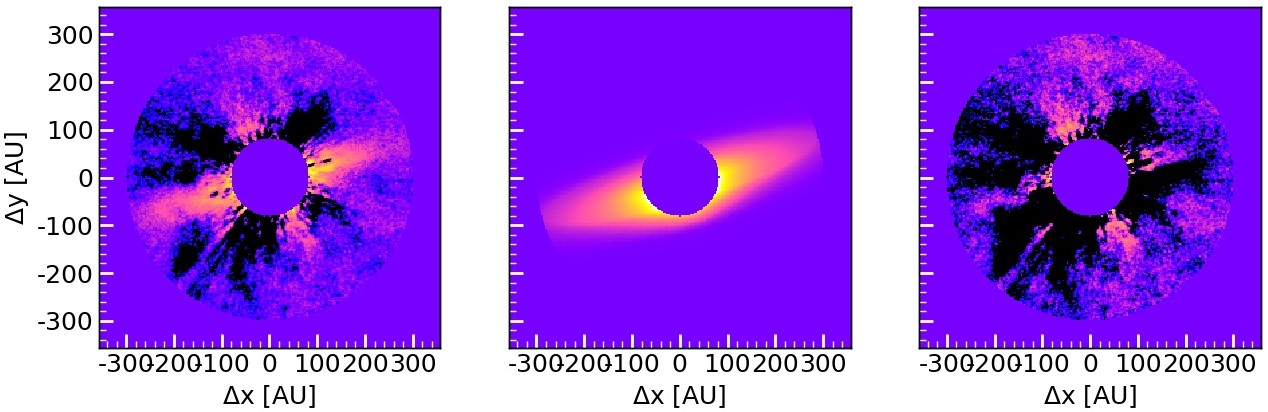}
\caption{Modelling results of 49~Cet. Left: ADI-reduced Y-band image as shown in Fig.~\ref{fig:ADI_image} excluding the region within 80~au. Middle: model generated with MODERATO assuming a single power law as radial distribution with $p=0.6$.  Right: Subtraction of the model disc from the observational data. The model leads to $\chi^2_\text{red} = 2.11$.}
\label{fig:ModelResults_SPHERE_1}
\end{figure*}

\begin{figure*}
\includegraphics[width=\textwidth]{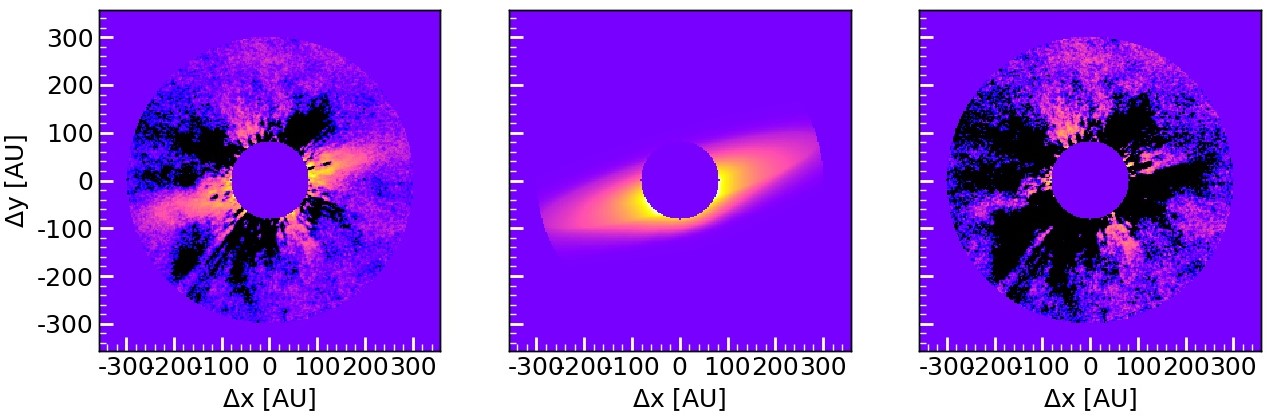}
\caption{Same as Fig.~\ref{fig:ModelResults_SPHERE_1}, but now using the single power law radial distribution with $p=0.5$ including an additional narrow ring at $r_\text{ring}=140~$au. For this model $\chi^2_\text{red} = 2.16$.}
\label{fig:ModelResults_SPHERE_2}
\end{figure*}

\begin{figure*}
\includegraphics[width=\textwidth]{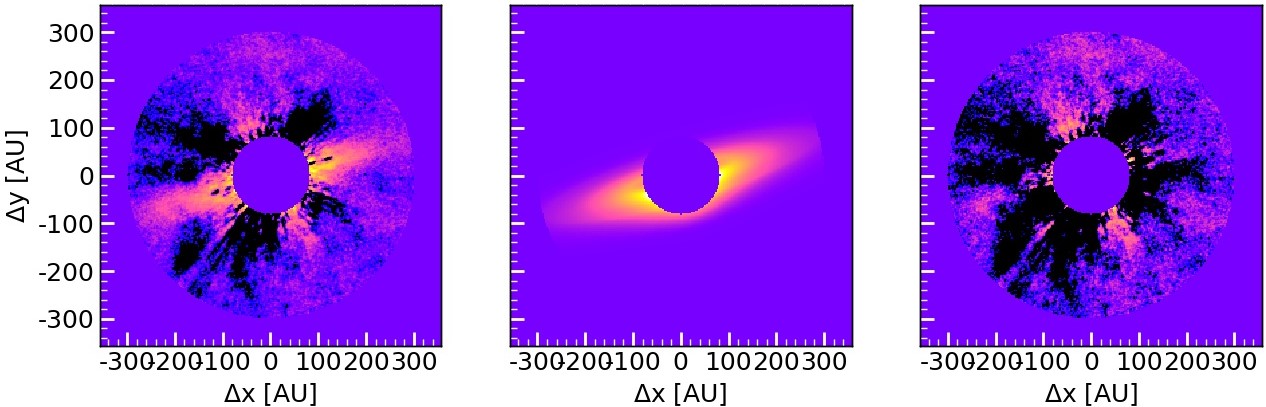}
\caption{
Same as Fig.~\ref{fig:ModelResults_SPHERE_1}, but now using the double power law radial profile with a transition radius at $r_\text{trans}=130~$au and radial distribution indices of $p_1 = -1.0$ and $p_2 = 1.0$. The best fit $\chi^2_\text{red}$ is 1.99. 
}
\label{fig:ModelResults_SPHERE_3}
\end{figure*}

In this section the best fitting models inferred from the ALMA data are compared to the scattered light data presented in this work. For reasons of comparability we also analyse the models with parameters stemming from \cite{hughes-et-al-2017}.
We use the grain size values given in Tab.~\ref{tab:grain_parameters} and the radial parameters listed in Tab.~\ref{tab:radial_parameters}.
The results are shown in Figs.~\ref{fig:ModelResults_SPHERE_1}-\ref{fig:ModelResults_SPHERE_3}.


The surface density distribution of each planetesimal belt is calculated in the same way as for the thermal emission maps. Each planetesimal then releases dust particles which are influenced by radiation pressure. From the position of the grain the appropriate scattered light is derived and summed up into a map. 
In contrast to the thermal emission images the scattered light maps possess a complex artefact structure inclined by 45$^\circ$and $90^\circ$ to the major axis which makes the interpretation of the modelling results difficult. So far there is no criterion to distinguish the artefact areas from the actual scattered light so that we cannot exclude these parts from $\chi^2$ calculations.

Nevertheless, we find that all three surface density profiles can describe the scattered light observations to a certain extent, but 
we see residuals along the major-axis and over-subtraction in the inner region close to the central star for all three radial profiles. However, the residuals in the outer disc region are at the same order as the estimated noise level of the image. 

Due to the artefact structures in the image in combination with the method to find the best fit (see Eq.~\ref{eq:chi2}), the $\chi2$ values inferred by forward modelling are very large and thus, we use again the $BIC$ introduced in the former section to compare the models with each other. The results are listed in Tab.~\ref{tab:ScatteredLight_parameters}.

\begin{table}
\caption{Comparison of scattered light models
\label{tab:ScatteredLight_parameters}}
\tabcolsep 3pt
\centering
\begin{tabular}{crr}
Comparing models &$\chi^2$ [$\times10^9$]& $B$		\\
\toprule
1, 2 & 8.58 & 4.8\\
1, 3 & 8.58 & 6.2\\
2, 3 & 8.58 & 4.8\\
\end{tabular}

\noindent
{\em Notes:}
The model numbers are given in Tab.~\ref{tab:ThermalEmission_parameters}.
The number of data points assumed is the number of pixels of the scattered light image: $N = 255 \times 255$. 
\end{table}

By following the classification of \citep{kass-raftery-1995} again, we find that the probability of a double power law profile is higher than the single power law and the power law including a narrow ring. Hence, the scattered light data lead to similar results as the thermal emission data showing that a double power law profile as depicted in Fig.~\ref{fig:SurfaceDensity} is more likely to occur in the debris disc around 49~Cet than the other profiles investigated in this study.

\section{Discussion}
\label{sec:discussion}
\subsection{Radial profiles}

We found that the radial extent of the 49~Ceti debris disc is the same for both ALMA and SPHERE observations as shown in Fig.~\ref{fig:Radial_Profile_ALMA}.
As theory predicts, the orbits of small grains should be highly altered by radiation pressure and thus, we would expect a larger disc extent for our scattered light image than for the thermal emission map. 
A reason could be the rather noisy scattered light detection of the disc only showing the brightest parts of it so that we underestimate its outer radius in scattered light. 

In comparison to other disc detections with SPHERE \citep[e.g., ][]{olofsson-et-al-2016, wahhaj-et-al-2016, engler-et-al-2019} the disc of 49~Cet while possessing a large fractional luminosity is rather faint in its surface brightness.  On the one hand this might indicate a dust composition strongly absorbing, such as carbon. On the other hand it might indicate the absence of small dust particles which is supported by the results of the SED fit and other studies, focusing on thermal emission and reporting that the disc is faint even in the mid-infrared \citep{wahhaj-et-al-2007}.
Another explanation is that the disc possesses a broad extent from at least 80~au up to 280~au (see Fig.~\ref{fig:Radial_Profile}).
Thus, a high number of small dust grains might be present in the disc, but due to its large on-sky area the particles are widely distributed so that the surface brightness stays low. In addition to that, the ADI reduction process \citep{marois-et-al-2006} leads to a stronger subtraction for broad discs and so the low detection might be an observational bias. 
 
 While the single power law profile was already excluded as the best-fit model by the aforementioned ALMA study \citep{hughes-et-al-2017}, the power law including a ring and the double power law profile remained indistinguishable for the analysis of thermal emission. To a certain extent we could reproduce this result with our semi-dynamical models, but we found that all three profiles are capable of describing the observational data (Fig.~\ref{fig:Radial_Profile_ALMA}).  Nevertheless, we found different parameter values for the best fitting fiducial discs. Applying the Bayesian information criterion we found that despite the comparable $\chi^2$ values for all three profiles, the double power law has the highest probability. 

Summarising the results we infer the following architecture of the outer region of the debris disc around 49~Cet.
From observed surface brightness profiles we find that the Kuiper-belt analogue is located between 80~au and 280~au and shows an increase of small particles between 130~au and 170~au visible in scattered light (Fig~\ref{fig:Radial_Profile}). Assuming the best-fitting double power law model, a maximum for large grains, seen at sub-mm wavelengths, is found around 110~au. The edges inwards and outwards of the ring are shallow. 
 
\subsection{Size distribution}
In \cite{pawellek-et-al-2014} the authors found a weak relation between the minimum grain size and the stellar luminosity of the host star. However, they stated that this trend is also consistent with being independent of the stellar luminosity and found an average value of roughly 5~$\mu$m to be valid for the majority of debris discs investigated. 
This is in good agreement with our SED-fit results where a value of $5.14\pm0.76\mum$ was found as dominant grain radius. The ratio of the minimum particle size to the inferred blow-out limit of 2.9$\mum$ for astrosilicate is 1.7. While it is directly connected to the dynamical excitation of dust-producing planetesimals in the disc, the low value of the ratio argues for a collisionally active disc for which a value around 2 is suggested \citep[e.g.,][]{krivov-et-al-2006, thebault-augereau-2007}. 

As shown in former studies \citep[e.g.,][]{pawellek-2017}, the size distribution index, $q$, is mostly influencing the steepness of the SED longwards of the far-infrared wavelengths. Hence, the now available VLA data at 9~mm \citep{macgregor-et-al-2016} help to strengthen the constraint. 
Compared to an idealised collisional cascade, where the index lies at $q=3.5$ \citep{dohnanyi-1969}, we inferred a value of $3.77\pm0.05$. This is somewhat larger, but still in agreement with collisional models \citep[e.g., ][]{loehne-et-al-2007, gaspar-et-al-2012, kral-et-al-2013, loehne-et-al-2017}.
The result supports the assumption of a dynamically excited disc producing more grains closer to the blow-out limit. 

Besides a higher excitation level which is expected for discs around earlier-type stars \citep{pawellek-krivov-2015}, the disc around 49~Cet is known to host a high amount of gas \citep[e.g.,][Moor  et al., submitted]{zuckerman-et-al-1995, hughes-et-al-2017}. The presence of gas might lead to a longer residence time of sub blow-out grains produced by collisions so that we would expect an increased number of small grains \citep{kral-et-al-2018}. 
However, we found the surface brightness to be rather weak in scattered light speaking against this assumption. 

A reason for a possible lack of small particles might be given by the blow-out limit for pure astronomical silicate, showing that grains smaller than 3$\mum$ are expelled from the 49~Cet system \citep{burns-et-al-1979, bohren-huffman-1983}. Fig.~\ref{fig:efficiencies} shows that grains smaller than 1$\mum$ possess the highest scattering efficiency. For larger grains this parameter stays nearly constant and thus, the scattered light is mainly determined by the particle cross section which is decreasing with increasing particle size, assuming a power law size distribution \citep[e.g.,][]{dohnanyi-1969}.
Furthermore, considering the inclined disc of 49~Cet, grains larger than 3.5$\mum$ are not as effectively contributing to the scattered light as smaller particles at the observational wavelength used \citep{zubko-2013}. 

Another possibility for the lack of small grains is a dynamically ``cold'' disc where an imbalance between the particles' production and destruction rate leads to a natural depletion of grains up to a few times the blow-out size \citep[see Fig.~7 of][]{thebault-wu-2008}. 
Furthermore small-grain depletion might be caused by the surface energy conservation criterion suggested by \cite{krijt-kama-2014}. Here, the minimum collisional fragment size is determined by the energy necessary to form the small grains. 
Nevertheless, the aforementioned mechanisms seem to be negligible for discs around A-type host stars \citep{thebault-2016}.

Thus, the explanation of a low surface brightness due to a broad on-sky area of the disc might be more appropriate than a lack of small grains although further studies are needed to confirm this suspicion.

\subsection{Model images}

\subsubsection{Thermal emission}
Considering the thermal emission maps we found elongated residuals in the outer disc region for the single power law profile including a narrow ring and the double power law profile which can be explained in different ways.
Firstly, they might be modelling artefacts since for the single power law model without a ring these extended residuals are not visible.
Secondly, the observations might suggest that there is a higher amount of grains in several parts of the outer disc than provided by the models. We do not see a symmetric distribution such as ring shaped residuals and thus, it is possible that the higher amount of particles is caused by recent collisions between massive bodies \citep{olofsson-et-al-2016} or stirring effects, such as self-stirring \citep[e.g.,][]{wyatt-2008, kennedy-wyatt-2010} or planetary stirring \citep[e.g.,][]{mustill-wyatt-2009}, although there were no planets found in the 49~Cet system so far \citep{choquet-et-al-2017}.

Besides the dust stirring mechanisms the particles might be dragged outward by present gas \citep[e.g.,][Mo\'or et al., submitted]{takeuchi-artymowicz-2001, thebault-augereau-2005, krivov-et-al-2009, kral-et-al-2018}. Nevertheless, the influence of the gas on dust particles may be limited to the smallest grains ($<5\mum$) not sensitively traced at sub-mm wavelengths. 

\subsubsection{Scattered light}
Comparing the scattered light maps to our ADI-reduced image we found that the double power law model reproduces the observations best.
However, residuals along the semi-major axis can be found in all residual images, but due to the complex artefact structure the remaining scattered light is of the order of the noise level that was estimated for the disc. 

In all semi-dynamical models applied to the scattered light observations, the grain size distribution inferred from SED fitting was used. 
Here, a minimum grain size of 5.14$\mum$ was assumed while the blow out grain size for 49~Cet lies at 2.9$\mum$ using pure astronomical silicate. The grains between 2.9$\mum$ and 5.14$\mum$ move on bound orbits around their host star and would therefore significantly contribute to the flux density in scattered light if they were present in the disc. 

\begin{figure*}
\includegraphics[width=\textwidth]{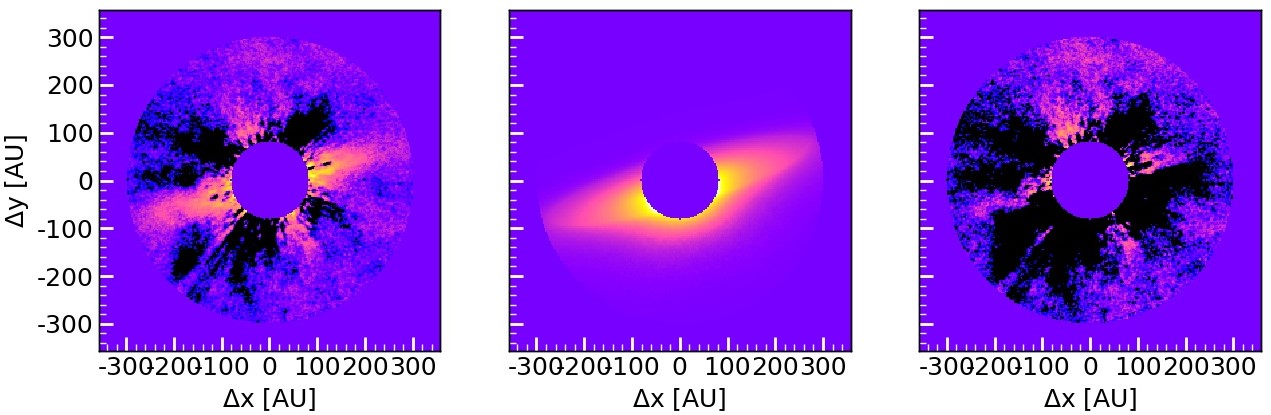}
\caption{
Same as Fig.~\ref{fig:ModelResults_SPHERE_1}, using the double power law radial profile and a size distribution between 2.9$\mum$ and 15$\mum$. The radial distribution index is $p=0.6$.
}
\label{fig:ModelResults_SPHERE_6}
\end{figure*}

Thus, we generated an additional model assuming the best fitting parameters of the single power law radial profile ($p=0.6$) and a size distribution ranging from the blow out grain size of 2.9$\mum$ to the upper cut-off size of 15$\mum$. 
The result can be seen in Fig.~\ref{fig:ModelResults_SPHERE_6}.

As suspected, the highly eccentric grains lead to a broader dust distribution and thus, the flux density in the outer disc regions increases comparable to the model using $5.14\mum$ as minimum grain size. 
Furthermore, the forward scattering becomes stronger so that the disc signal is overestimated close to the star leading to an increased $\chi^2_\text{red}$ of $2.27$ in comparison to the models using the SED size distribution where we found $\chi^2_\text{red}=2.11$.
In an extended study we analysed scattered light models using a grid of different minimum grain sizes between the blow-out limit (2.9$\mum$) and the best-fit size of the SED (5.14$\mum$). The model using a size of 5$\mum$ delivers the best result with $\chi^2_\text{red}=2.11$. Thus, the semi-dynamical models give us an additional opportunity to estimate the minimum dust grain size besides a pure SED fit.

\subsection{Scattering properties and impact on our results}

An explanation for the residuals, although being close to the estimated noise level, might be given by the simplified model assumption considering the dust grains to be compact spheres in order to use Mie-theory. It is well established that the optical properties, such as absorption and scattering efficiencies or asymmetry parameter, strongly depend on the particle shape which can lead to significant deviations from the Mie values assumed \citep[e.g., ][]{schuerman-et-al-1981, weiss-wrana-1983, mugnai-et-al-1986}. Hence, in comparison to non-spherical grains, the Mie theory approach overestimates the fraction of forward scattering for larger particles \citep[e.g.,][]{arnold-et-al-2018}. 
In order to stay in an appropriate grain size regime, we estimated an upper cut-off grain size of $15\mum$ for particles still significantly contributing to the measured scattered light and omitted all larger particles \citep[see Sec.~\ref{sec:MODERATO}, ][]{zubko-2013}. 

To weaken the effect of forward scattering, many debris disc studies use the Henyey-Greenstein approach \citep{henyey-greenstein-1941}  where the asymmetry parameter, $g$, is fixed to a specific value between -1 and 1 to give an integrated scattering phase function independent of particle composition and sizes. 

\begin{figure*}
\includegraphics[width=\textwidth]{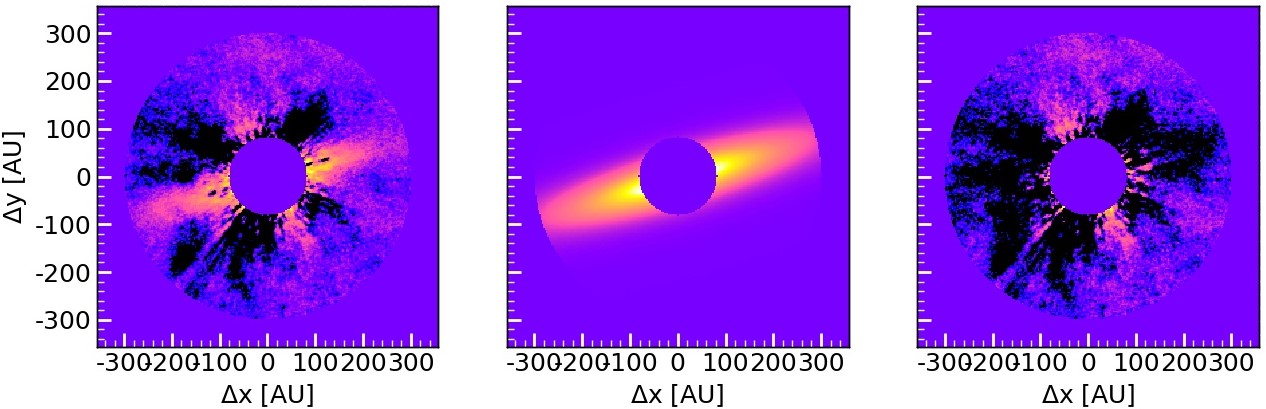}
\caption{
Same as Fig.~\ref{fig:ModelResults_SPHERE_1}, using the single power law radial profile with a fixed asymmetry parameter of $g=0.1$. 
}
\label{fig:ModelResults_SPHERE_5}
\end{figure*}

In the first scattered light study of 49~Cet, \cite{choquet-et-al-2016} gave a best fit value of $g=0.1$. 
To check whether the remaining emission is caused by the overestimation of forward scattering we calculated a set of models based on the single power law radial distribution and fixed the asymmetry parameter for each fiducial disc. In this approach, the $g$ parameter is varied between -0.9 and 0.9 in steps of 0.1. We found a best fit model with $g=0.1$ which is in agreement with \cite{choquet-et-al-2017}. The result is shown in Fig.~\ref{fig:ModelResults_SPHERE_5}.

Since the amount of forward scattering is reduced in the Henyey-Greenstein approach, we could reduce the overestimation of scattered light close to the star. 
Nevertheless, due to the higher particle eccentricities (see Appendix~\ref{sec:comparison} for more details) in comparison to the discs assuming Mie theory, we get an overestimation of flux density in the outer disc region by assuming the radial distribution index to be $p = 0.6$.

At this point, we have to give a word of caution about the usage of the Henyey-Greenstein approach together with our semi-dynamical dust modelling since it is an approach often used in former studies \citep[e.g.,][]{Lee-Chiang-2016, esposito-et-al-2016, olofsson-et-al-2016}.
The MODERATO code uses Mie theory to calculate the optical parameters of the generated dust grains. However, it also serves as basis to compute the radiation pressure parameter, $\beta$ (see Eq.~\ref{eq:beta}), and therefore the dust particle orbits. It could be shown that $\beta$ depends as strongly on the shape of the dust grains as the optical parameters and thus, can easily differ by 75\% comparing equivalent surface areas of non-spherical and spherical grains \citep[e.g.,][]{schuerman-et-al-1981}. 
By fixing the asymmetry parameter, the optical properties of the dust material assumed are tampered and hence, using the Henyey-Greenstein approach is inconsistent with the dust population inferred from Mie theory. 
Furthermore, to calculate the flux density of the scattered light not only the asymmetry parameter, but the scattering and extinction efficiencies are needed. These parameters are not provided by the Henyey-Greenstein approach and thus, have to be taken from Mie calculations or comparable methods which enlarges the already existing inconsistency. A more detailed comparison between the Henyey-Greenstein approach and Mie theory can be found in Appendix~\ref{sec:comparison}.

In contrast to Mie and Henyey-Greenstein theory, a more sophisticated approach to model the scattered light might be given in the Discrete Dipole Approximation \citep[DDA, e.g.,][]{draine-flatau-2013}.

\section{Summary}
\label{sec:summary}

We performed a multi-wavelength study of the debris disc around 49~Cet focusing on the dust component. We presented new scattered light data obtained with the SPHERE/IRDIS instrument on the VLT in H23 dual band and Y band filters and used thermal emission data from a former ALMA study at $850\mum$. 
The H23-band detection was weak in comparison to the Y-band data so we focused our analysis on the Y-band data. For future SPHERE observations of comparable debris discs we therefore suggest using a broad band filter for stronger detections. 

In both wavelength regimes the disc radius was found to be $\sim280$~au. While for the ALMA detection a clear decrease in surface brightness could be observed, the extent of the disc in scattered light is limited by the noise level and might therefore be underestimated.

In a first modelling step, we fitted the SED of 49~Cet to obtain information on the grain size distribution. We inferred a dominant particle radius of 5.14$\mum$ and a size distribution index of 3.77, suggesting that the disc is dynamically excited. 

We computed semi-dynamical models using three different radial surface density profiles. By comparing our models to the observations in thermal emission and scattered light we found that all three can reproduce both SPHERE and ALMA observations, but that the double power law profile achieves the best fit results. 
Hence, we assume that the small grains ($\sim5\mum$) dominant in scattered light and large particles dominant in thermal emission stem from the same planetesimal belt and are mainly influenced by radiation pressure and collisions.
Since transport processes, such as Poynting-Robertson drag, have greater effect on small particles we would expect differences in the radial dust distribution of small and large grains. However, these differences might only be observable in the inner disc region between the planetesimal belt and the star \citep{kennedy-piette-2015}. Since our observations are not sensitive for this inner region of the 49~Cet debris disc we cannot draw any conclusions on such inward directed transport processes.

\section*{Acknowledgements}
NP is thankful for fruitful discussions with Torsten L\"ohne and Alexander Krivov.
AM acknowledges support from the Hungarian National Research, Development and Innovation Office NKFIH Grant KH-130526. 
JO acknowledges financial support from the ICM (Iniciativa Cient\'ifica Milenio) via the N\'ucleo Milenio de Formaci\'on Planetaria grant, from the Universidad de Valpara\'iso, and from Fondecyt (grant 1180395)
This project has received funding from the European Research Council (ERC) under the European Union's Horizon 2020 research and innovation programme under grant agreement No 716155 (SACCRED).
FMe acknowledges funding from ANR of France under contract number ANR-16-CE31-0013.
AZ acknowledges support from the CONICYT + PAI/ Convocatoria nacional subvenci\'on a la instalaci\'on en la academia, convocatoria 2017 + Folio PAI77170087. 
This work has made use of data from the European Space Agency (ESA) mission
{\it Gaia} (\url{https://www.cosmos.esa.int/gaia}), processed by the {\it Gaia}
Data Processing and Analysis Consortium (DPAC,
\url{https://www.cosmos.esa.int/web/gaia/dpac/consortium}). Funding for the DPAC
has been provided by national institutions, in particular the institutions
participating in the {\it Gaia} Multilateral Agreement.
SPHERE is an instrument designed and built by a consortium
consisting of IPAG (Grenoble, France), MPIA (Heidelberg, Germany), LAM
(Marseille, France), LESIA (Paris, France), Laboratoire Lagrange
(Nice, France), INAF–Osservatorio di Padova (Italy), Observatoire de
Genève (Switzerland), ETH Zurich (Switzerland), NOVA (Netherlands),
ONERA (France) and ASTRON (Netherlands) in collaboration with
ESO. SPHERE was funded by ESO, with additional contributions from CNRS
(France), MPIA (Germany), INAF (Italy), FINES (Switzerland) and NOVA
(Netherlands).  SPHERE also received funding from the European
Commission Sixth and Seventh Framework Programmes as part of the
Optical Infrared Coordination Network for Astronomy (OPTICON) under
grant number RII3-Ct-2004-001566 for FP6 (2004–2008), grant number
226604 for FP7 (2009–2012) and grant number 312430 for FP7
(2013–2016). We also acknowledge financial support from the Programme National de
Planétologie (PNP) and the Programme National de Physique Stellaire
(PNPS) of CNRS-INSU in France. This work has also been supported by a grant from
the French Labex OSUG@2020 (Investissements d’avenir – ANR10 LABX56).
The project is supported by CNRS, by the Agence Nationale de la
Recherche (ANR-14-CE33-0018). It has also been carried out within the frame of the National Centre for Competence in 
Research PlanetS supported by the Swiss National Science Foundation (SNSF). MRM, HMS, and SD are pleased 
to acknowledge this financial support of the SNSF.
Finally, this work has made use of the the SPHERE Data Centre, jointly operated by OSUG/IPAG (Grenoble), PYTHEAS/LAM/CESAM (Marseille), OCA/Lagrange (Nice) and Observatoire de Paris/LESIA (Paris) and is supported by a grant from 
Labex OSUG@2020 (Investissements d'avenir - ANR10 LABX56). 
We thank P. Delorme and E. Lagadec (SPHERE Data
Centre) for their efficient help during the data reduction
process.





\newcommand{\AAp}      {A\& A}
\newcommand{\AApR}     {Astron. Astrophys. Rev.}
\newcommand{\AApS}     {AApS}
\newcommand{\AApSS}    {AApSS}
\newcommand{\AApT}     {Astron. Astrophys. Trans.}
\newcommand{\AdvSR}    {Adv. Space Res.}
\newcommand{\AJ}       {AJ}
\newcommand{\AN}       {AN}
\newcommand{\AO}       {App. Optics}
\newcommand{\ApJ}      {ApJ}
\newcommand{\ApJL}     {ApJL}
\newcommand{\ApJS}     {ApJS}
\newcommand{\ApSS}     {Astrophys. Space Sci.}
\newcommand{\ARAA}     {ARA\& A}
\newcommand{\ARevEPS}  {Ann. Rev. Earth Planet. Sci.}
\newcommand{\BAAS}     {BAAS}
\newcommand{\CelMech}  {Celest. Mech. Dynam. Astron.}
\newcommand{\EMP}      {Earth, Moon and Planets}
\newcommand{\EPS}      {Earth, Planets and Space}
\newcommand{\GRL}      {Geophys. Res. Lett.}
\newcommand{\JGR}      {J. Geophys. Res.}
\newcommand{\JOSAA}    {J. Opt. Soc. Am. A}
\newcommand{\MemSAI}   {Mem. Societa Astronomica Italiana}
\newcommand{\MNRAS}    {MNRAS}
\newcommand{\PASJ}     {PASJ}
\newcommand{\PASP}     {PASP}
\newcommand{\PSS}      {Planet. Space Sci.}
\newcommand{\RAA}      {Research in Astron. Astrophys.}
\newcommand{\SolPhys}  {Sol. Phys.}
\newcommand{\SolSysRes}{Sol. Sys. Res.}
\newcommand{\SSR}      {Space Sci. Rev.}

\bibliographystyle{mnras}
\bibliography{FinalePaper.bbl}


\appendix

\section{Theoretical background}
\label{sec:theory}
The orbits of dust particles are altered by different mechanisms, such as collisions, Poynting-Robertson drag, and stellar radiation pressure is one of the strongest processes.

\subsection{Radiation pressure}
As shown in earlier studies \citep[e.g.,][]{burns-et-al-1979, wyatt-et-al-1999} the orbital parameters depend on the radiation pressure parameter, $\beta$, given as
\begin{equation}
	\beta \equiv \frac{\left|\vec{F_\text{rad}}\right|}{\left|\vec{F_\text{G}}\right|} = \frac{3L_\text{star}}{16\pi G c M_\text{star}} \frac{\overline{Q}_\text{pr}}{\varrho s}.
    \label{eq:beta}
\end{equation}
Here, $L_\text{star}$ and $M_\text{star}$ are the stellar luminosity and mass, $G$ the gravitational constant, $c$ the speed of light, $Q_\text{pr}$ the radiation pressure efficiency averaged over the stellar spectrum, $\varrho$ the bulk density of the dust material and $s$ the grain radius. 
If we assume a dust particle is released from a planetesimal possessing the orbital parameters semi-major axis, $a_\text{p}$, eccentricity, $e_\text{p}$ and true anomaly $f_\text{p}$, then the orbit parameters of the dust grain can be calculated with
\begin{equation}
	a_\text{d} = \frac{a_\text{p}(1-\beta)(1-e_\text{p}^2)}{1-e_\text{p}^2-2\beta(1+e_\text{p}\cos(f_\text{p}))}
    \label{semimajor_axis}
\end{equation}
\begin{equation}
	e_\text{d}^2 = \frac{\beta^2 + e_\text{p}^2+ 2\beta e_\text{p}\cos(f_\text{p})}{(1-\beta)^2}
    \label{eccentricity}
\end{equation}
\begin{equation}
	\tan(\varpi-\varpi_\text{p}) = \frac{\beta \sin(f_\text{p})}{\beta \cos(f_\text{p})+ e_\text{p}},
    \label{pericentre}
\end{equation}
where $\varpi$ is the longitude of pericentre. 
To compute the $\beta$ parameter, we calculate $Q_\text{pr}$ using Mie theory, where the particles are assumed to be compact spheres \citep{bohren-huffman-1983}.  

\begin{figure*}
\includegraphics[width=0.35\textwidth, angle=-90]{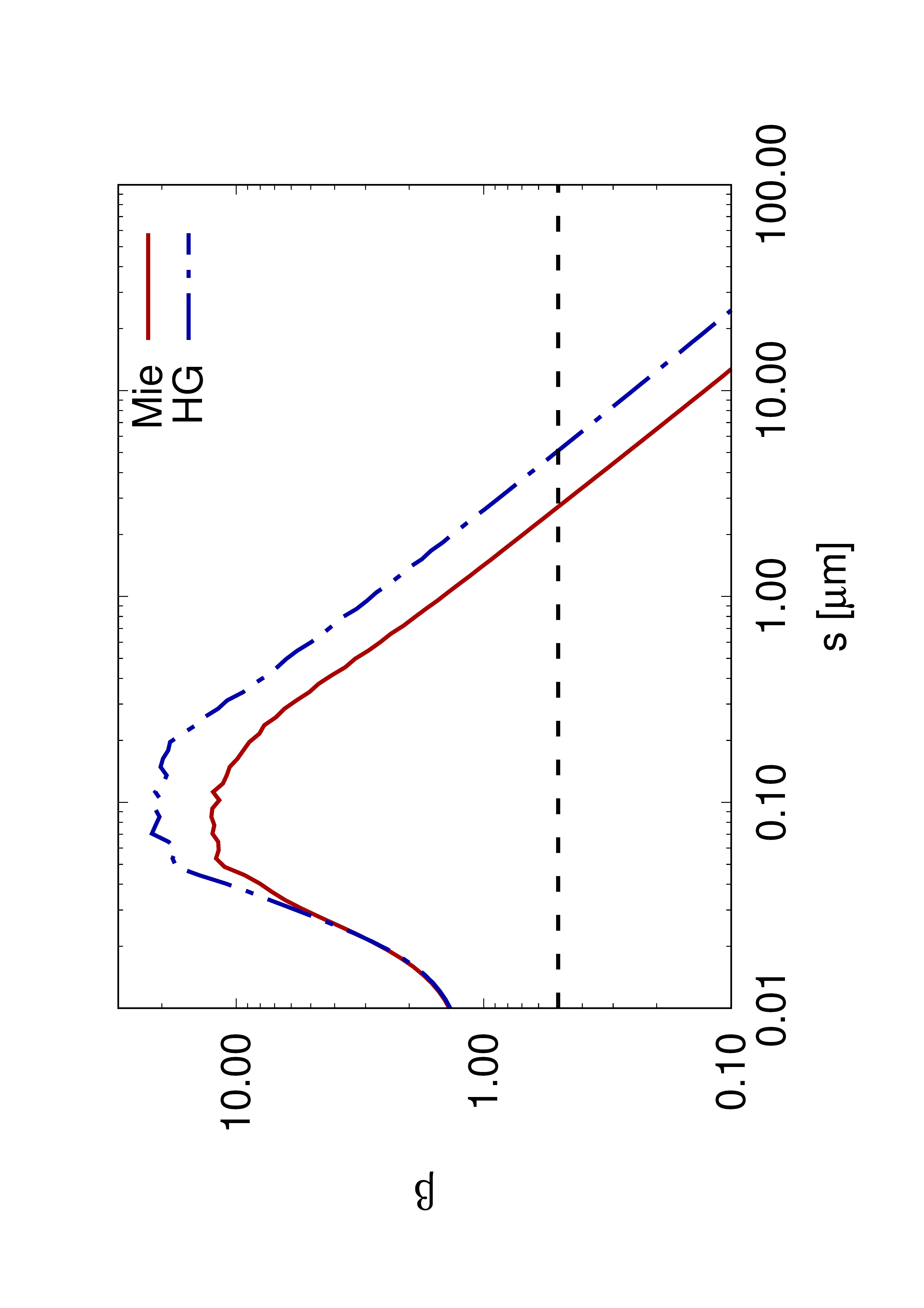}
\includegraphics[width=0.35\textwidth, angle=-90]{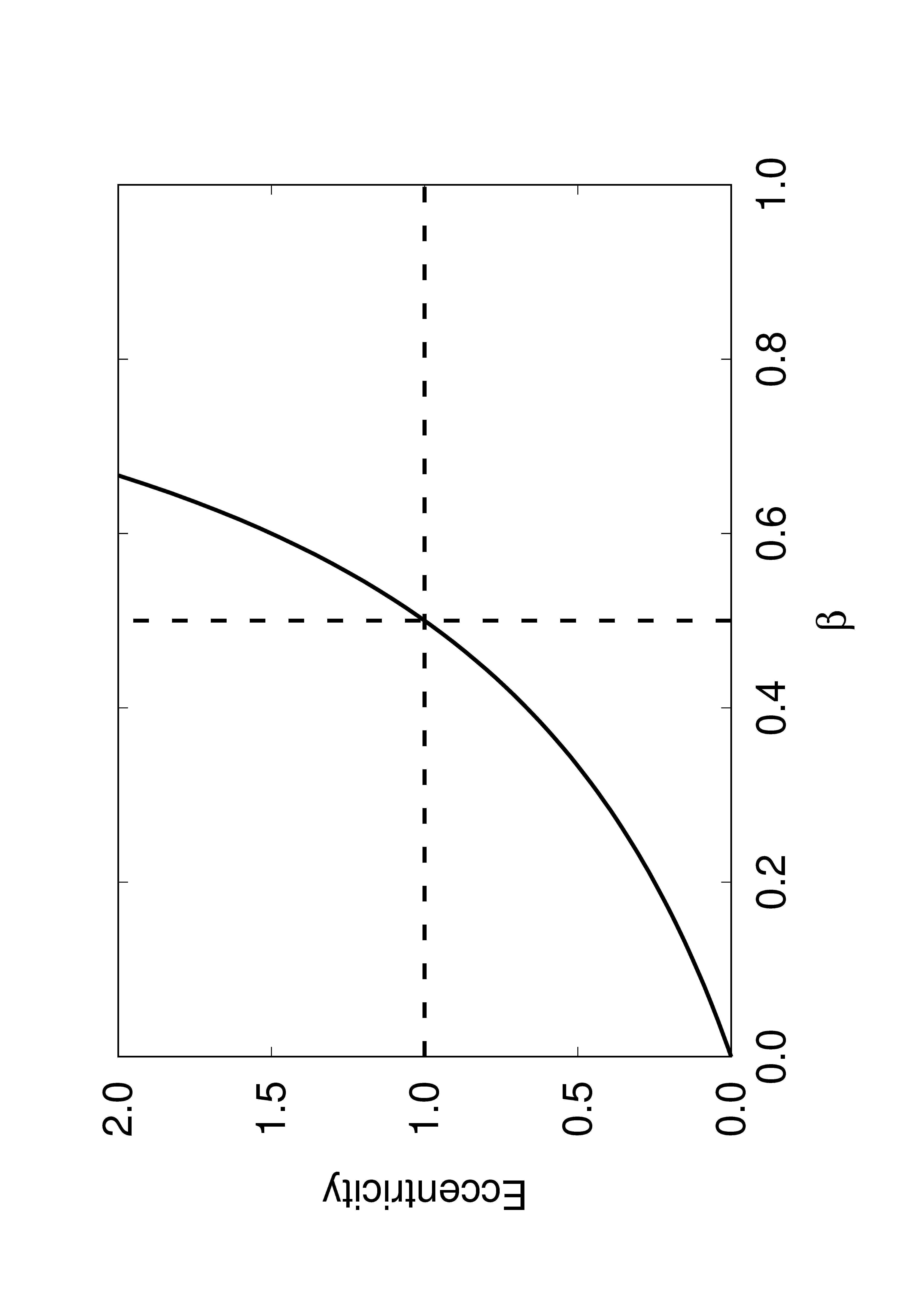}
\caption{Left: Radiation pressure parameter as a function of grain size calculated for 49~Cet (solid black line). Solid red line: using Mie theory; dash-dotted blue line: using Henyey-Greenstein approach with $g=0.1$; black dashed line shows $\beta=0.5$. Right: particle eccentricity as a function of the radiation pressure parameter (solid black line) assuming a circular orbit for the parent body. The dashed lines show $\beta=0.5$ and $e=1.0$.}
\label{fig:RadiationPressure}
\end{figure*}

In Fig.~\ref{fig:RadiationPressure} on the left side the $\beta$ parameter is shown as a function of grain size. To calculate $\beta$, we use the stellar properties of 49~Cet and astronomical silicate \citep{draine-2003} with a bulk density of 3.3~g/cm$^3$. 
On the right side the particle eccentricity is given as a function of $\beta$. We use Eq.~\ref{eccentricity} and assume $e_\text{p}$ and $f_\text{p}$ to be equal to zero. In this case, the eccentricity turns 1 when $\beta$ is 0.5 and thus, grains below a value of $\beta=0.5$ stay in bound orbits around the star, while for values larger than 0.5 the particles are expelled from the stellar system on either parabola or hyperbola orbits. 
In general, $\beta$ is increasing with decreasing grain size. Therefore, the so-called blow-out grain size for 49~Cet is 2.9$\mum$ meaning, that grains smaller than this value are expelled from the stellar system. 

Furthermore, the smaller the grains the larger the eccentricity gets. As a consequence, small grains spend more time in the apocentre region compared to large particles and thus, it is expected that by tracing small grains in observations we would see larger disc radii than by tracing large grains. 

To calculate the dust disc we generate at first the parent belt using the surface number density. In many cases a size independent power law is assumed:
\begin{equation}
    N(r) = N_0\times\left(\frac{r}{r_0}\right)^{-p},
\end{equation}
where the parameters are the same as described in Sec.~\ref{sec:SED}.
Now we have a population of planetesimals at different radii and angles releasing the dust grains. The planetesimals give us the parameters necessary ($a_\text{p}$, $e_\text{p}$, $f_\text{p}$) to calculate the particle orbits ($a_\text{d}$, $e_\text{d}$).

\subsection{Tracing different grain sizes}

\begin{figure}
\begin{center}
\includegraphics[width=0.8\columnwidth, angle=-90]{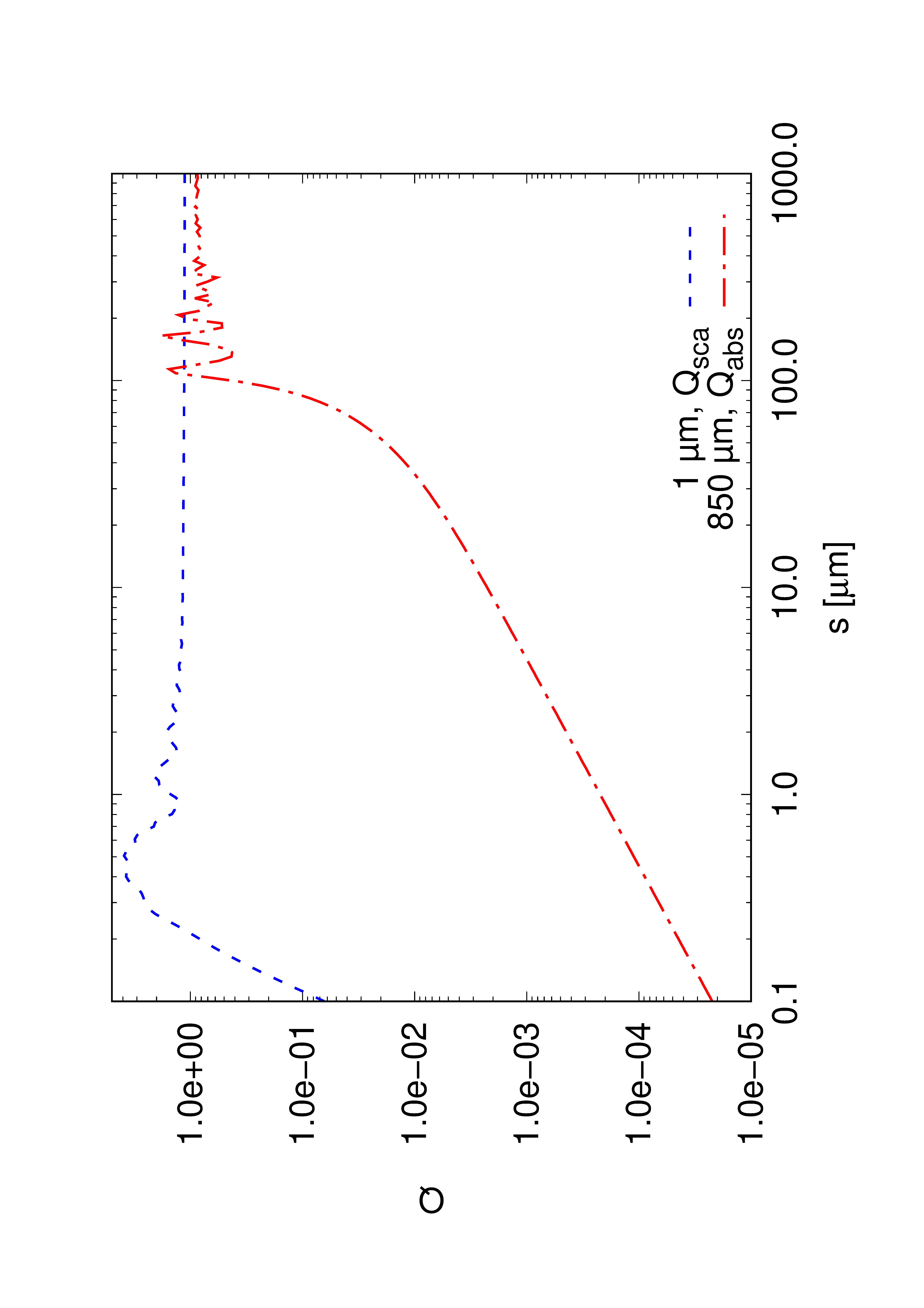}
\caption{Scattering and absorption efficiency as a function of grain size. The blue dashed line shows the scattering efficiency  at a wavelength of 1.04$\mum$ and the red dash dotted line the absorption efficiency at 850$\mum$. The dust composition is pure astronomical silicate.}
\label{fig:efficiencies}
\end{center}
\end{figure}

It is necessary to investigate the absorption and scattering efficiencies for each grain size and wavelength to get information on the emission of the particles. 
For thermal emission the absorption efficiency of the grains is of importance while for scattered light emission it is the scattering efficiency. 
In Fig.~\ref{fig:efficiencies} the absorption efficiency is depicted for a wavelength of 850$\mum$ and the scattering efficiency for 1.04$\mum$. These are the wavelengths used by SPHERE and ALMA for the 49~Cet observations used in this study.
Grains with a size of $\sim0.5\mum$ posses the highest scattering efficiency while $\sim200\mum$ sized particles show the largest absorption efficiency at the observed wavelengths. This means, we would expect to trace the small grains best with scattered light emission in the near infrared and the large grains with thermal emission in the sub-mm region. However, taking the radiation pressure into account, we expect to see grains larger than the blow-out size. 

\begin{figure*}
\includegraphics[width=\textwidth]{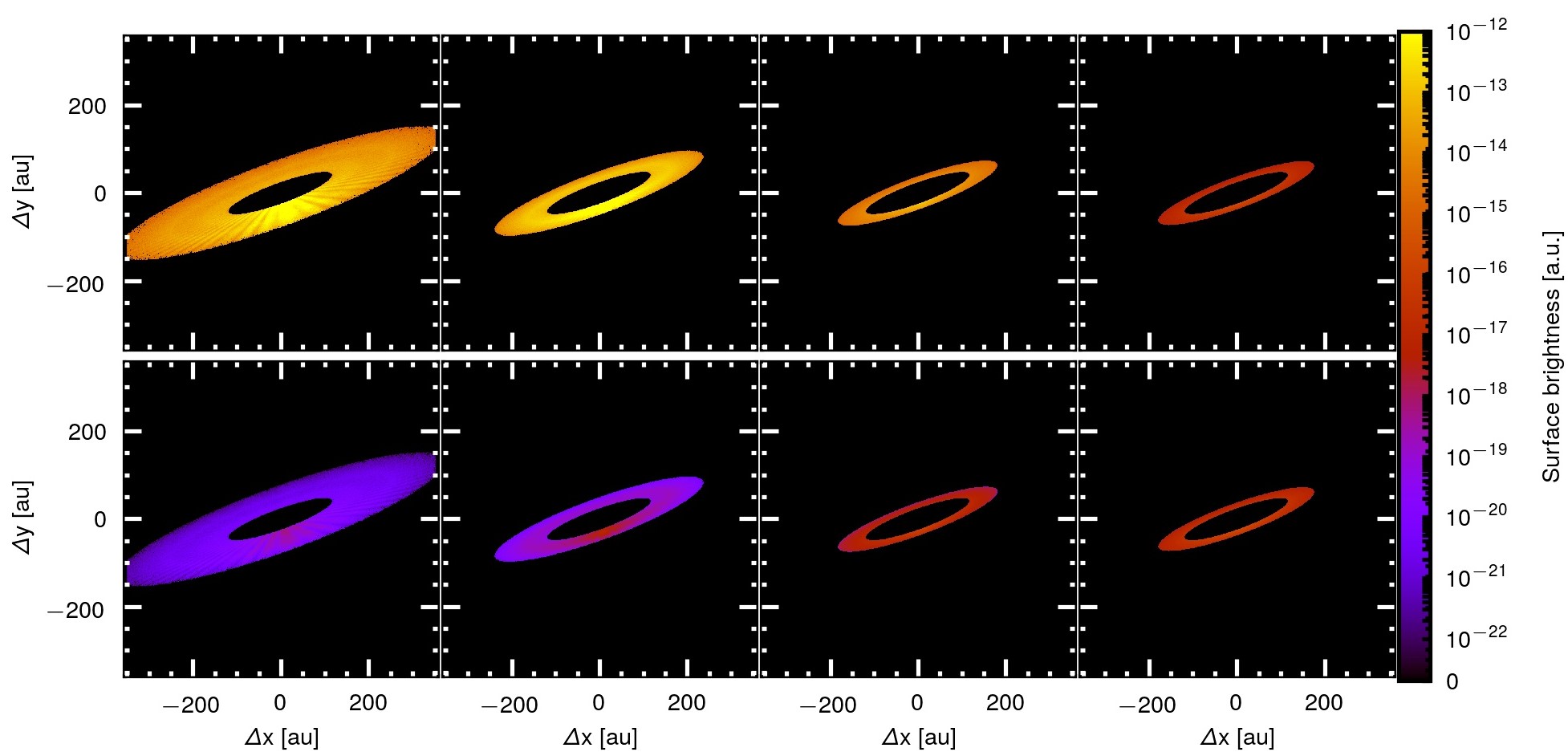}\\
\includegraphics[width=\textwidth]{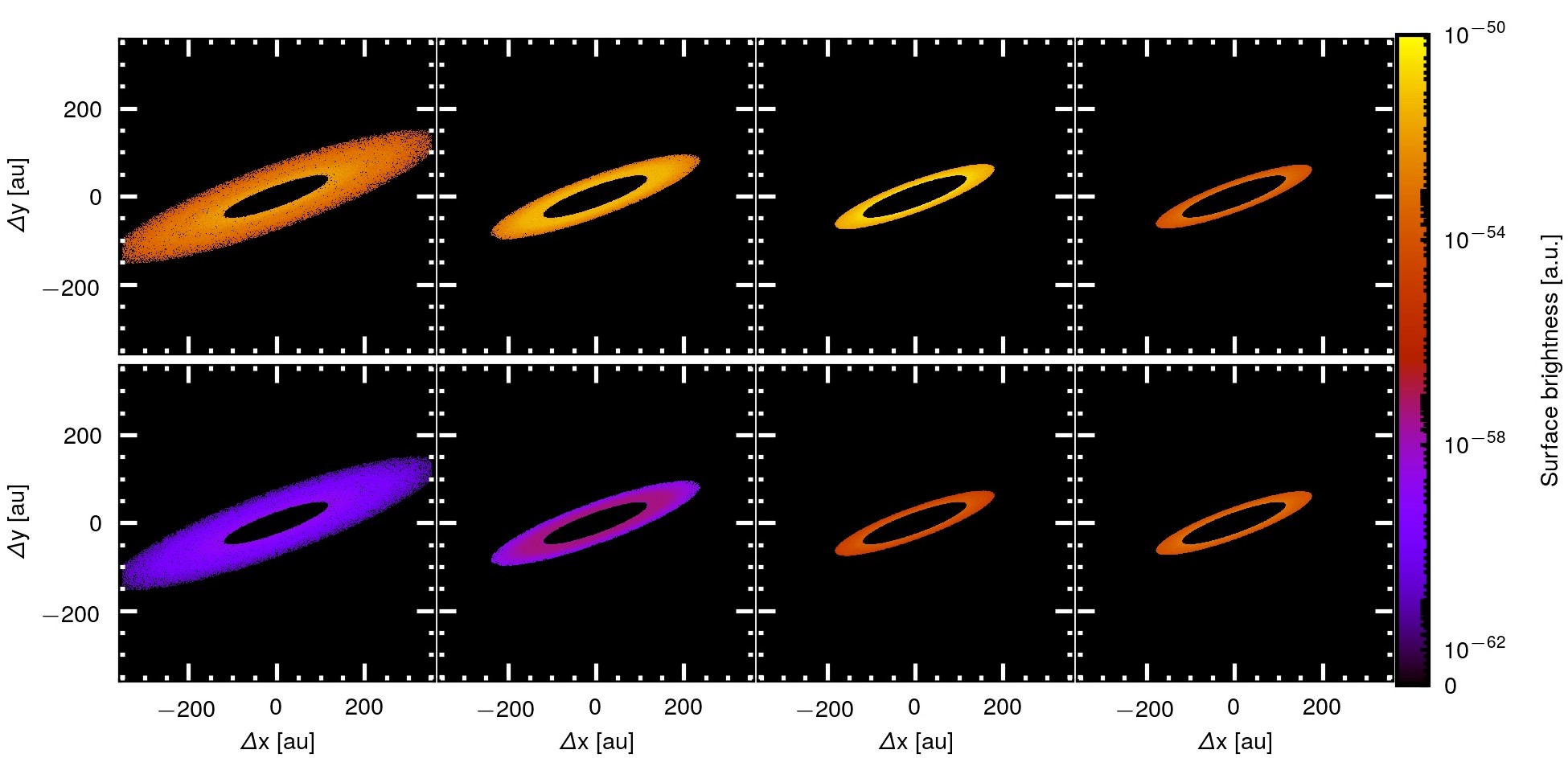}
\caption{Model discs for different grain sizes. From left to right: The dust is made of a single grain size of 5$\mum$, 10$\mum$, 100$\mum$, 1000$\mum$. The first row gives the scattering emission of the discs of a constant mass at 1.04$\mum$. The second row shows the same as the first row, but now the discs have the same number of particles. The third row shows the discs with a constant mass in thermal emission at 850$\mum$. The bottom row shows the same as the third row, but now the discs have a constant number of particles.}
\label{fig:modeldiscs}
\end{figure*}

In Fig.~\ref{fig:modeldiscs} fiducial discs are given for thermal emission and scattered light. We use the 1.04$\mum$ and 850$\mum$ as observational wavelengths. Furthermore, we assume two cases. The first being that all discs possess the same mass and the second that all discs contain the same number of particles.
The scattered light images at 1.04$\mum$ are shown in the first two rows, where the first row represents discs of the same mass and the second row discs with the same particle number. 
We see in the first case that the discs get fainter for an increasing grain size and decrease in radius. The latter fact is mainly caused by stellar radiation pressure.
In the second case, where the particle number stays constant the discs get brighter. 

One reason is the decreasing scattering efficiency. However, it stays nearly the same for grains larger than 10$\mum$ and thus, the second reason is the constant disc mass leading to a smaller number of large grains. Another point is the wavy structure of the emission is caused by the phase function which depends on the grain size, the observation wavelength and the scattering angle. 
Considering the thermal emission images it is obvious that the disc with dust made of 100$\mum$ grains possesses the largest emission at 850$\mum$ compared to 5, 10 and 1000$\mum$. The absorption efficiency decreases for particles smaller than 100$\mum$. Hence, even the larger number of small particles cannot compensate the lower absorption.

\section{Comparison of Mie theory and Henyey-Greenstein approach}
\label{sec:comparison}

As shown by Equation~\ref{eq:beta}, the radiation pressure parameter, $\beta$, is directly proportional to the radiation pressure efficiency, $\overline {Q}_\text{pr}(s)$, averaged over the stellar spectrum which is given by
\begin{equation}
\overline{Q}_\text{pr} = \frac{\int Q_\text{pr}(s, \lambda)\times F_\lambda \text{d}\lambda}{\int F_\lambda\text{d}\lambda}.
\end{equation}
Here, $Q_\text{pr}$ describes the radiation pressure efficiency depending on grain radius, $s$, and wavelength, $\lambda$, while $F_\lambda$ gives the stellar flux density.
For each grain and wavelength, $Q_\text{pr}$ is calculated by
\begin{equation}
Q_\text{pr}(s, \lambda) = Q_\text{ext}(s, \lambda) - Q_\text{sca}(s, \lambda)\times \langle \cos(\vartheta)\rangle(s, \lambda),
\end{equation}
where $Q_\text{ext}$ and $Q_\text{sca}$ are the extinction and scattering efficiencies and $\langle\cos(\vartheta)\rangle$ the asymmetry parameter, depending on the scattering angle, $\vartheta$ \citep{bohren-huffman-1983}. This parameter is also called $g$. All the parameters used to calculate $Q_\text{pr}$ depend on the grain size and the wavelength. 

Using the Henyey-Greenstein approach (HG), the asymmetry parameter is fixed and therefore independent of grain size and wavelength, while all other parameters are still calculated by Mie theory or comparable methods.
Since the extinction and scattering efficiencies as well as the $g$ parameter depend on the optical constants, i.e. refractive indices, of the dust composition used, the fixation of $g$ resembles a variation of the dust material. 

\begin{figure}
\includegraphics[width=0.8\columnwidth, angle = -90]{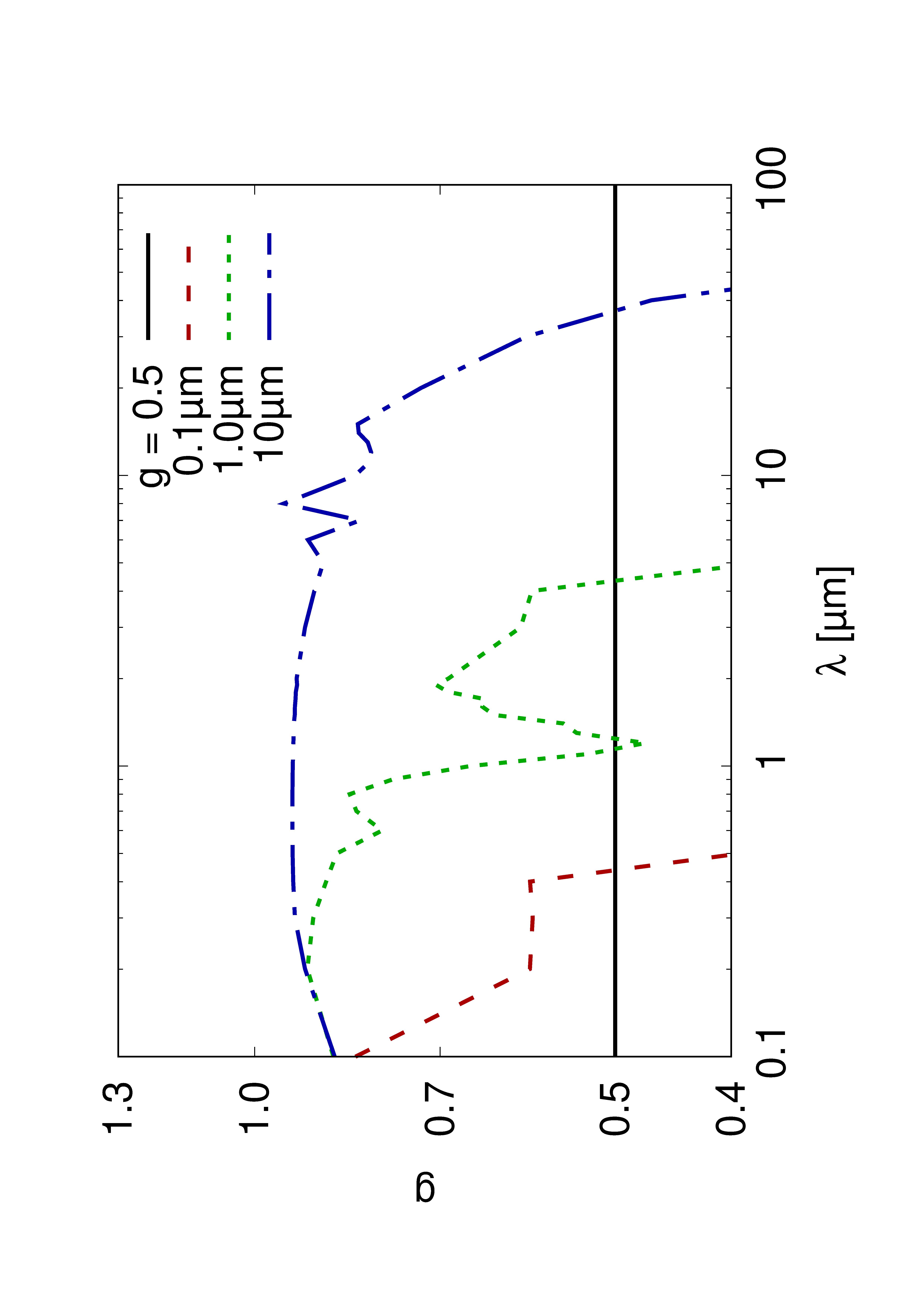}
\caption{Asymmetry parameter as function of wavelength. Comparison of Mie theory and Henyey-Greenstein approach using astronomical silicate and a fixed $g$ of 0.5. Black solid line shows the fixed asymmetry parameter; colour dashed lines show g for different grain radii (red dashed: 0.1$\mum$, green dotted: 1.0$\mum$, blue dash-dotted: 10$\mum$). }
\label{fig:HG_1}
\end{figure}

A comparison of $g$ values given by Mie and HG is presented in Fig.~\ref{fig:HG_1}. Here, $g$ is set to 0.5 resembling a best-fit value for different studies using the HG approach \citep[e.g.,][]{millar-blanchaer-et-al-2015, olofsson-et-al-2016, engler-et-al-2017, engler-et-al-2019}.
The values from Mie show strong deviations from this value. 
For 0.1$\mum$-sized particles, $g$ decreases to values close to zero for wavelength larger than $1\mum$ while for larger grains this decrease is moved to longer wavelengths. 

\begin{figure}
\includegraphics[width=0.8\columnwidth, angle = -90]{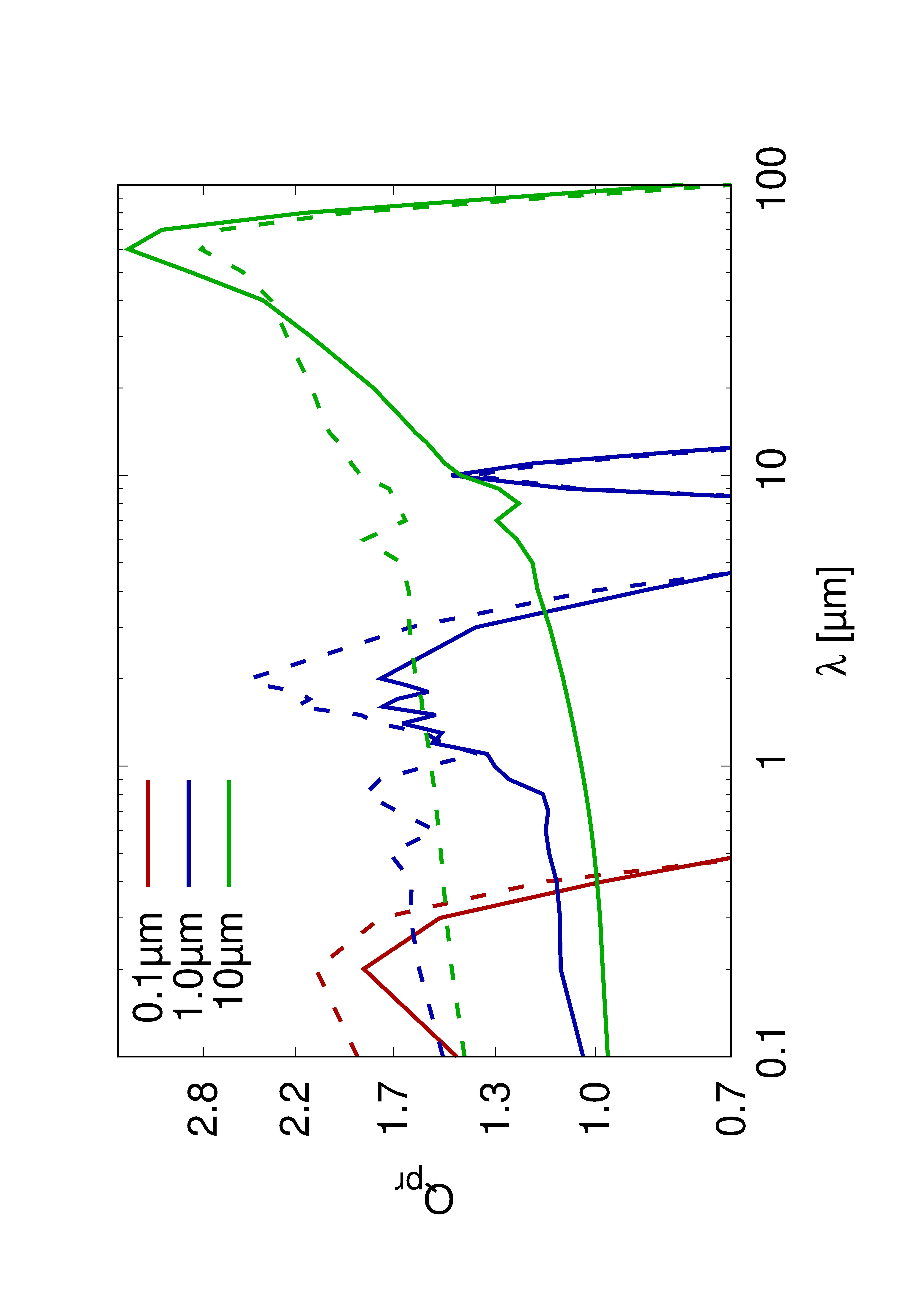}
\caption{Radiation pressure efficiency as function of wavelength for the same grain radii as in Fig.~\ref{fig:HG_1}. The solid lines shows the result for Mie theory, the dashed lines for Henyey-Greenstein assuming $g=0.5$. }
\label{fig:HG_2}
\end{figure}

In Fig.~\ref{fig:HG_2} we present the radiation pressure efficiency as a function of wavelength for different grain radii. Depending on the particle size, HG and Mie lead to similar results for wavelengths roughly longer than $\sim3\times s$. For shorter wavelengths, the differences between the methods can easily reach 50\%, although these deviations decrease for smaller particles (assuming $g=0.5$). 

\begin{figure}
\includegraphics[width=0.8\columnwidth, angle = -90]{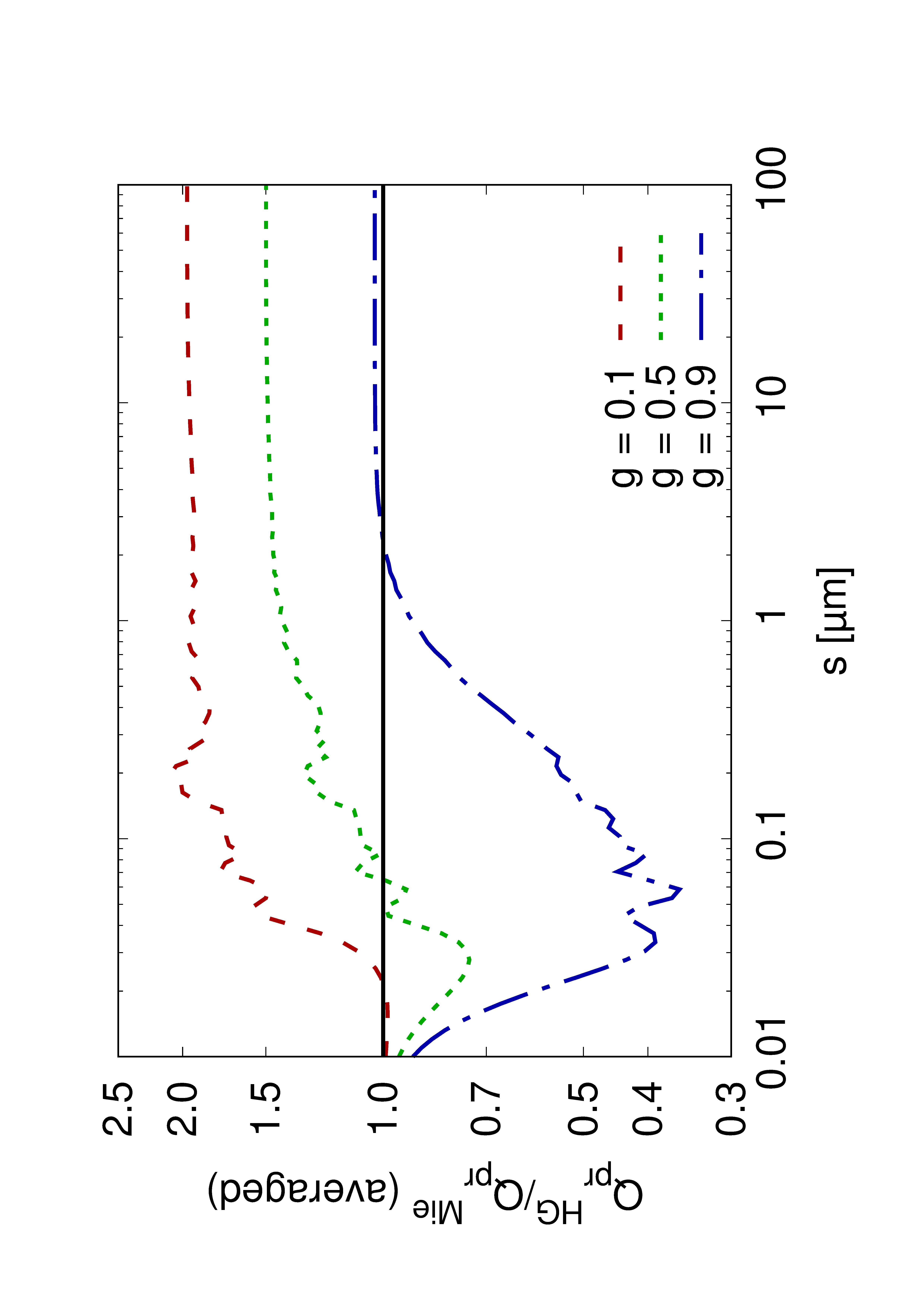}
\caption{Ratio of the averaged radiation pressure efficiencies of Henyey-Greenstein and Mie theory as a function of grain size for different values of g. Red dashed: $g=0.1$; green dotted: $g=0.5$, blue dash-dotted: $g=0.9$; black solid: ratio of 1}
\label{fig:HG_3}
\end{figure}

In the next step we compare the radiation pressure efficiencies averaged over the stellar spectrum for both, HG and Mie (Fig.~\ref{fig:HG_3}). Here, the value calculated by HG is divided by the value inferred with Mie theory and given as a function of grain radius. Using three different $g$ values between 0.1 and 0.9 for HG, $\overline{Q}_\text{pr}$ is overestimated for both, $g=0.1$ and 0.5 with differences of up to $\sim100\%$ assuming $g=0.1$ and $\sim$50\% assuming $g=0.5$. 
In case of $g=0.9$, the deviation of $\overline{Q}_\text{pr}$ is close to zero for grains larger than $\sim2\mum$ while we underestimate  $\overline{Q}_\text{pr}$ for smaller particles. 

The parameter $\beta$ changes by the same order of magnitude as $\overline{Q}_\text{pr}$ (see Fig.~\ref{fig:RadiationPressure}). 
Hence, similar-sized particles possess different eccentricities for both methods . For example, if we assume a grain with a certain size $s$ for which $\beta=0.1$ using Mie theory, we get $\beta$ for HG twice as large assuming $g=0.1$ (see Figs. \ref{fig:RadiationPressure}, \ref{fig:HG_3}). This leads to an increase of the particle's eccentricity by a factor of 2.25 compared to Mie, using Eq.~\ref{eccentricity} and a circular parent belt. For larger $\beta$ this effect is even stronger due to the non-linear relation between $\beta$ and grain eccentricity (see Fig.~\ref{fig:RadiationPressure}). 
Only for the smallest particles ($s \sim 0.01\mum$) HG and Mie reach similar values. 
Hence, by applying the Henyey-Greenstein approach the dust population inferred with Mie theory is altered by HG.


\bsp	
\label{lastpage}
\end{document}